\documentclass[12pt]{article}
\usepackage{amsmath}

\numberwithin{equation}{section}

\begin{document}
\baselineskip=17pt
\begin{titlepage}
\begin{flushright}
{\footnotesize TUM-HEP-608/05}\\[-1mm]
{\footnotesize KYUSHU-HET-79}\\[-1mm]
{\small hep-ph/0511108}
\end{flushright}
\begin{center}
\vspace*{8mm}

{\large\bf%
Discrete Flavor Symmetry, Dynamical Mass Textures,\\[1mm]
and Grand Unification%
}\vspace*{7mm}

Naoyuki Haba$^{\rm a}\,$ and Koichi Yoshioka$^{\rm b}$
\vspace*{3mm}

{\it $^{\rm a}$Physik-Department, Technische Universit\"at M\"unchen,}\\
{\it James-Franck-Stra{\ss}e, 85748 Garching, Germany}\\[1mm]
{\it $^{\rm b}$Department of Physics, Kyushu University, Fukuoka,
812-8581, Japan} 

\vspace*{3mm}

{\small (November, 2005)}
\end{center}
\vspace*{5mm}

\begin{abstract}\noindent%
Discrete flavor symmetry is explored for an intrinsic property of
mass matrix forms of quarks and leptons. In this paper we
investigate the $S_3$ permutation symmetry and derive the general
forms of mass matrices in various types of $S_3$ theories. We also 
exhibit particular realizations of previous ansatze of mass matrices,
which have often been applied in the literature to the standard model
Yukawa sector. Discrete flavor symmetry is also advantageous for
vanishing matrix elements being dynamically generated in the vacuum of
scalar potential. This is due to the fact that group operations are
discrete. While zero elements themselves do not explain mass
hierarchies, we introduce an abelian flavor symmetry. A non-trivial
issue is whether successful quantum numbers can be assigned so that
they are compatible with other (non-abelian) flavor symmetries. We
show typical examples of charge assignments which not only produce
hierarchical orders of mass eigenvalues but also prohibit
non-renormalizable operators which disturb the hierarchies in
first-order estimation. As an explicit application, a flavor model is
constructed in grand unification scheme 
with $S_3$ and $U(1)$ (or $Z_N$) flavor symmetries.
\end{abstract}
\footnote[0]{{\footnotesize \hspace*{-5mm}$^{\rm a\,}$On leave of
absence from Institute of Theoretical Physics, University of
Tokushima, 770-8502, Japan.}}
\end{titlepage}

%%%%%%%%%%%%%%%%%%%%%%%%%%%%%%%%%%%%%%%%%%%%%%%%%%%%%%%%%%%%%%%%%%%%%%
\section{Introduction}

One of the most important issues in and beyond the standard model is
the masses and mixing angles of the three-family quarks and
leptons. After the electroweak symmetry breaking, the observed values
of masses and mixing angles are to be produced from the structures of
Yukawa couplings. However the Yukawa couplings generally have
redundancy in explaining the experimental results: apparently
different forms of Yukawa matrices lead to the same physical masses
and mixing angles. Therefore it has been a long outstanding problem
which patterns of Yukawa couplings are relevant from phenomenological
and theoretical viewpoints. Various progresses have been made in the
literature by applying additional principles to the standard-model
Yukawa sector. The two well-known examples of such principles are to
adopt the unification hypothesis of matter multiplets and to assume
specific forms of Yukawa couplings (``textures''). The former is a
top-down approach to the problem. The grand unification principle
relates the properties of quarks and leptons, and reduces the degrees
of freedom of Yukawa couplings in the theory. On the other hand, the
latter approach is rather a bottom-up one. Available forms of Yukawa
textures are explored so that they are consistent with the
experimental observations. It is interesting that the number of
successful textures is found to be highly limited. This fact is
revealed from a simplifying assumption that some elements of
Yukawa matrices vanish, called texture zeros. Along this line,
phenomenologically possible forms of mass matrices have been proposed
in the literature~(for example,~\cite{Fritzsch,GJ,Giudice}, and
also~\cite{RRR,FGM} for systematic analyses of zero textures).

As for the neutrino sector, the recent experimental results suggest
that one of the most likely forms of Majorana mass textures of light
neutrinos is proportional to
\begin{equation}
  \begin{pmatrix}
    ~~~~~& & \\ & {\cal O}(1) & \!{\cal O}(1) \\ 
    & {\cal O}(1) & \!{\cal O}(1)
  \end{pmatrix}
  \label{1111}
\end{equation}
in the basis where the generation mixing has been rotated out in the
charged-lepton side. This is the dominant part of mass matrix and
other small entries have not explicitly been written down. It is
clearly seen that the large leptonic mixing between the second and
third generations~\cite{atmos} is 
predicted from (\ref{1111}). The observed large 1-2
mixing~\cite{solar} requires an additional condition for the above
texture: the dominant $2\times2$ sub-matrix of (\ref{1111}) has
a reduced rank and its determinant is of the same order of small
off-diagonal elements neglected in (\ref{1111}). Given that, the
texture form (\ref{1111}) is considered as a promising candidate
consistent to the present experimental data. The condition (the
vanishing determinant of the dominant sub-matrix) may be realized
without fine tuning of model parameters, e.g., with the 
right-handed neutrino dominance~\cite{sneudom}, $R$-parity
violation~\cite{Rparity}, and the lopsided form of charged-lepton mass
matrix~\cite{lopsided}. It is interesting that the lopsided mass
textures can be naturally embedded in grand unified 
theory. However if the theory is supersymmetrized, large off-diagonal
elements in the lepton Yukawa matrices generally induce sizable rates
of flavor-violating processes to excess the present experimental
bounds~\cite{STY}. While the details depend on superparticle mass
spectrum, a natural way to avoid this flavor problem is to consider
the case that the lepton as well as quark Yukawa matrices take
hierarchical forms, which lead to small generation mixing. If this is
the case, an interesting possibility to have large lepton mixing is to
suppose asymmetric forms of mass 
textures (zeros).\footnote{Leptonic mixing angles may be 
enhanced, e.g.\ by integrating out heavy fields such as right-handed
neutrinos~\cite{enhance}.} It seems that abelian flavor symmetry may
be difficult to generate such asymmetric zeros without expense of
model complexity. Furthermore non-abelian continuous flavor symmetry
is not suitable for handling mass textures since texture zeros are
rotated to other arbitrary forms by continuous symmetry rotations and
do not have physical implications.

Motivated by these results, in this paper we investigate the power of
non-abelian discrete flavor symmetry for constructing mass matrix
models. We particularly focus on the minimal discrete non-abelian 
symmetry $S_3$. (For fermion mass models based on other non-minimal
discrete non-abelian flavor 
symmetries, see~\cite{flavor}.) \ The $S_3$ operation is the
permutation of three objects, which has a simple geometrical 
interpretation, i.e.\ the symmetry of an equilateral triangle. While
it is the smallest non-abelian discrete symmetry, it might be regarded
as a remnant of flavor symmetry of fundamental theory in high-energy
regime. The purposes of this paper are the following two 
points: (i) non-abelian flavor symmetries such as $S_3$ are so
effective that various types of phenomenological textures are handled
and (ii) these symmetries can also generate asymmetric forms of Yukawa
matrices in dynamical ways. Further it should be noted that texture
zeros themselves do not explain fermion mass hierarchies. Previous
approaches to fermion masses with $S_3$ have assumed hierarchical
values of Yukawa couplings and/or involved symmetry-breaking patterns
such as vacuum expectation values (VEVs) of Higgs scalars. In this
paper we show that realistic values of masses and mixing angles are
dynamically achieved by introducing $U(1)$ symmetry. A non-trivial
problem arises whether successful $U(1)$ quantum numbers can be
assigned so that they are compatible with other 
non-abelian (flavor) symmetries. It is noticed that mass hierarchy is
also realized in a similar way with a discrete subgroup of the 
flavor $U(1)$ symmetry such as $Z_N$ with 
appropriate (enough large) $N$ and the same quantum numbers as in 
the $U(1)$ case. A smaller choice of $N$ would be possible and
interesting from a viewpoint of brevity. In this case, the problem of
fermion masses can be handled with flavor symmetries that are entirely
discrete. While the $U(1)$ charge assignments are presented in this
paper, they can always be read as the charges in 
flavor $Z_N$ theory. We finally present an explicit model where 
the $S_3$ flavor symmetry is incorporated consistently to unified
gauge symmetry and hierarchical forms of mass matrices.

This paper is structured as follows. In the next section, we discuss
some fundamental issues of the $S_3$ group, which are needed in
Section 3 to study symmetry-invariant forms of mass matrices. In
Section 4, $S_3$ is applied to supply dynamical justifications to
realistic candidates of mass textures which have often been discussed
in the literature. Based on these results, we present in Section 5 a
toy flavor model in grand unification scheme where Yukawa textures are
controlled by a single flavor $S_3$, assisted by $U(1)$ symmetry. In
Section 6, we analyze the invariant scalar potentials 
of $S_3$ doublet whose VEV form is a key ingredient of the approach 
developed in this paper. Section 7 is devoted to summarizing our results.

%%%%%%%%%%%%%%%%%%%%%%%%%%%%%%%%%%%%%%%%%%%%%%%%%%%%%%%%%%%%%%%%%%%%%%
\section{The $S_3$ group}
\label{S3group}
%%%%%%%%%%%%%%%%%%%%%%%%%%%%%%%%%%%%%%%%%%%%%%%%%%%%%%%%%%%%%%%%%%%%%%
\subsection{Representations and representation matrices}

The $S_3$ symmetry is the smallest non-abelian symmetry, the
permutations which an equilateral triangle has. The $S_3$ group
therefore contains six elements $T_1$, $\cdots$, $T_6$, half of which
are the circulations of triangle apices and the other half corresponds
to the exchanges of two of three apices while the other is
fixed. There are only few numbers of irreducible representations; a
two-dimensional representation and two different one-dimensional
representations. Throughout of this paper we denote them 
as $2$ (doublet), $1_{\rm S}$ (singlet), and $1_{\rm A}$ (pseudo
singlet), respectively. The non-trivial one-dimensional 
representation $1_{\rm A}$ distinguishes the group elements in two
parts. The matrix representations of group elements are given in
Table~\ref{matrix}. 
\begin{table}[htbp]
\begin{center}
\begin{tabular}{c|c|c|c|c|c|c}
& $T_1$ & $T_2$ & $T_3$ & $T_4$ & $T_5$ & $T_6$ \\ \hline
2 & $\begin{pmatrix}1 & 0 \\ 0 & 1\end{pmatrix}$ & 
$\begin{pmatrix}\chi & 0 \\ 0 & \chi^2\end{pmatrix}$ &
$\begin{pmatrix}\chi^2 & 0 \\ 0 & \chi\end{pmatrix}$ &
$\begin{pmatrix}0 & 1 \\ 1 & 0\end{pmatrix}$ &
$\begin{pmatrix}0 & \chi^2 \\ \chi & 0\end{pmatrix}$ &
$\begin{pmatrix}0 & \chi \\ \chi^2 & 0\end{pmatrix}$ \\ \hline
$1_{\rm S}$ & 1 & 1 & 1 & 1 & 1 & 1 \\ \hline
$1_{\rm A}$ & 1 & 1 & 1 & $-1$ & $-1$ & $-1$
\end{tabular}
\end{center}
\caption{The representation matrices of the $S_3$ elements. The 
symbol $\chi$ is the third root of unity $(\chi=e^{2\pi i/3})$.}
\label{matrix}
\end{table}
It is easily seen that the even 
permutations $T_1$, $T_2$ and $T_3$ constitute the 
subgroup $Z_3$. This means that the $S_3$ symmetry is broken down 
to $Z_3$ when a pseudo-singlet field $1_{\rm A}$ develops an
expectation value. Thus the $S_3$ group has simple but non-trivial
structures, and is suitable for applying it to the flavor problems of
three-generation fermions in the standard model. A number of models
have been proposed to explain Yukawa coupling structures of quarks and
leptons with the $S_3$ flavor symmetry~\cite{S3demo,S3real,S3cpx}.

It may be convenient to introduce the reducible three-dimensional
representations for discussing the three-generation flavor
physics. Corresponding to $1_{\rm S}$ and $1_{\rm A}$, there are two
types of three-dimensional representations:
\begin{eqnarray}
3_{\rm S} &\equiv& 2+1_{\rm S}, \quad (\textrm{triplet})\\
3_{\rm A} &\equiv& 2+1_{\rm A}. \quad (\textrm{pseudo triplet})
\end{eqnarray}
If $S_3$ is regarded as a remnant of some gauge symmetry in
fundamental theory, $3_{\rm A}$ should be applied not to induce
discrete gauge anomaly. This is understood from the fact that the
three-dimensional vector representation of $SO(3)$ is decomposed 
as $3=2+1_{\rm A}$ in terms of its subgroup $S_3$. One may also use
the $3_{\rm S}$ representation at the expense that $S_3$ is assumed to
be a global symmetry or the anomaly is cancelled by introducing
appropriate numbers of pseudo singlet fermions. Further the
aforementioned geometrical interpretation of $S_3$ operations is made
clear for triplet representations. Such an interpretation is seen in a
different basis of $S_3$, as will be discussed later in this
section. The matrix representations 
of $3_{\rm S}$ and $3_{\rm A}$ are read from Table~\ref{matrix} and
given by
\begin{alignat}{6}
T_1 &= \begin{pmatrix}
  1 & & \\ & 1 & \\ & & 1
\end{pmatrix},& \qquad
T_2 &= \begin{pmatrix}
  \chi & & \\ & \chi^2 & \\ & & 1
\end{pmatrix},& \qquad
T_3 &= \begin{pmatrix}
  \chi^2 & & \\ & \!\chi\, & \\ & & 1
\end{pmatrix}, \nonumber \\
T_4 &= \begin{pmatrix}
  & 1 & \\ 1 & & \\ & & 1
\end{pmatrix},& \qquad
T_5 &= \begin{pmatrix}
  & \chi^2 & \\ \chi & & \\ & & 1
\end{pmatrix},& \qquad
T_6 &= \begin{pmatrix}
  & \!\chi\, & \\ \chi^2 & & \\ & & 1
\end{pmatrix},
\label{Tcpx}
\end{alignat}
for the $3_{\rm S}$ representation. As for $3_{\rm A}$, the 
matrices $T_i$ are given by (\ref{Tcpx}) except that the 3-3 elements
in the odd permutation matrices $T_{4,5,6}$ are replaced 
with $-1$.

%%%%%%%%%%%%%%%%%%%%%%%%%%%%%%%%%%%%%%%%%%%%%%%%%%%%%%%%%%%%%%%%%%%%%%
\subsection{Tensor products and $\boldsymbol{2^*}$ representation}

The tensor products involving $1_{\rm A}$ are given 
by $1_{\rm A}\times 1_{\rm A}=1_{\rm S}$ 
and $1_{\rm A}\times 2=2$. The only remaining non-trivial product is
that of two doublets: $2\times 2=2+1_{\rm A}+1_{\rm S}$. In the basis
where the group elements are given by Table~\ref{matrix}, the product
of two doublets $\psi=(\psi_1,\psi_2)^{\rm t}$ 
and $\phi=(\phi_1,\phi_2)^{\rm t}$ is explicitly written as follows:
\begin{eqnarray}
  \psi \times \phi &=& (\psi^\dagger\sigma_+\phi,\,
  \psi^\dagger\sigma_-\phi)_2^{\rm t} 
  +(\psi^\dagger\sigma_3\phi)_{1_{\rm A}}
  +(\psi^\dagger\phi)_{1_{\rm S}}, \label{22} \\
  &=& (\psi_1^\dagger\phi_2,\,\psi_2^\dagger\phi_1)_2^{\rm t}
  +(\psi_1^\dagger\phi_1-\psi_2^\dagger\phi_2)_{1_{\rm A}}
  +(\psi_1^\dagger\phi_1+\psi_2^\dagger\phi_2)_{1_{\rm S}},
\end{eqnarray}
where $\sigma_i$ ($i=1,2,3$) are the Pauli matrices 
and $\sigma_\pm\equiv\frac{\sigma_1\pm i\sigma_2}{2}$. The subscripts
in the right-handed sides denote $S_3$ representations. Note 
that $\psi$ (and also $\phi$) is generically complex-valued, 
while $S_3$ is a real group. This fact is important 
when $\psi$ transforms as some complex representation under other
symmetries than $S_3$. The complex conjugate $\psi^*$ belongs to 
the $2^*$ representation for which the representation matrices 
become $T_i^*$. In practical use, however, it is convenient to
define a doublet from an anti-doublet:
\begin{equation}
  \psi_{{}_C} \,\equiv\, \sigma_1\psi^* \,=\, 
  \begin{pmatrix} \psi_2^* \\ \psi_1^* \end{pmatrix},
\end{equation}
which transforms as $\psi_{{}_C}\to\sigma_1(T_i\psi)^*=
T_i\sigma_1\psi^*=T_i\psi_{{}_C}$, and indeed acts 
as the $2$ representation of $S_3$. The tensor product 
involving $\psi_{{}_C}$ is hence given by
\begin{eqnarray}
  \psi_{{}_C}\!\times \phi &=& (\psi^{\rm t}\sigma_L\phi,\,
  \psi^{\rm t}\sigma_R\phi)_2^{\rm t} 
  +(\psi^{\rm t} i\sigma_2\phi)_{1_{\rm A}}
  +(\psi^{\rm t}\sigma_1\phi)_{1_{\rm S}}, \label{22c} \\
  &=& (\psi_2\phi_2,\,\psi_1\phi_1)_2^{\rm t}
  +(\psi_1\phi_2-\psi_2\phi_1)_{1_{\rm A}}
  +(\psi_1\phi_2+\psi_2\phi_1)_{1_{\rm S}},
\end{eqnarray}
where $\sigma_L\equiv\frac{1-\sigma_3}{2}$ 
and $\sigma_R\equiv\frac{1+\sigma_3}{2}$, respectively. It is noticed
that this product does not contain any complex conjugates. Such a type
of tensor product is necessary for describing, e.g., matter
interaction terms ($F$ terms) in supersymmetric theory and Majorana
masses for neutrinos. On the other hand, the product (\ref{22}) is
applied to the usual Dirac mass terms of quarks and leptons. The form
of tensor product depends on the group basis, and therefore 
the $S_3$ model construction and its physical consequences also
do. For details, see the next subsection and Section~\ref{inv}.

It may be useful for later discussion to explicitly write down what
types of singlet components are contained in the products of more than
two doublets. In the basis discussed here, one finds
\begin{alignat}{3}
  &\psi\times\phi &\;\;\;\supset\;\;\;
  &\psi_1^\dagger\phi_1+\psi_2^\dagger\phi_2,\quad
  \psi_1\phi_2+\psi_2\phi_1, \\
  &\psi\times\phi\times\varphi &\supset\;\;\;
  &\psi_1^\dagger\phi_2\varphi_2+\psi_2^\dagger\phi_1\varphi_1,\quad
  \psi_1\phi_2^\dagger\varphi_1+\psi_2\phi_1^\dagger\varphi_2,
  \nonumber \\
  &&& \psi_1\phi_1\varphi_2^\dagger+\psi_2\phi_2\varphi_1^\dagger,
  \quad \psi_1\phi_1\varphi_1+\psi_2\phi_2\varphi_2, \\
  &\psi\times\phi\times\varphi\times\chi &\supset\;\;\;
  &\psi_1^\dagger\phi_1^\dagger\varphi_1\chi_1
  +\psi_2^\dagger\phi_2^\dagger\varphi_2\chi_2,\quad
  \psi_1\phi_1^\dagger\varphi_1^\dagger\chi_1
  +\psi_2\phi_2^\dagger\varphi_2^\dagger\chi_2, \nonumber \\
  &&& \psi_1\phi_1^\dagger\varphi_1\chi_1^\dagger
  +\psi_2\phi_2^\dagger\varphi_2\chi_2^\dagger, \quad
  \psi_1^\dagger\phi_1\varphi_1\chi_2
  +\psi_2^\dagger\phi_2\varphi_2\chi_1, \nonumber \\
  &&& \psi_1\phi_1^\dagger\varphi_1\chi_2
  +\psi_2\phi_2^\dagger\varphi_2\chi_1, \quad
  \psi_1\phi_1\varphi_1^\dagger\chi_2
  +\psi_2\phi_2\varphi_2^\dagger\chi_1, \nonumber \\
  &&& \psi_1\phi_2\varphi_2\chi_2^\dagger
  +\psi_2\phi_1\varphi_1\chi_1^\dagger, \quad
  \psi_1^\dagger\phi_2\varphi_1\chi_1
  +\psi_2^\dagger\phi_1\varphi_2\chi_2, \nonumber \\
  &&& \psi_1^\dagger\phi_1\varphi_2\chi_1
  +\psi_2^\dagger\phi_2\varphi_1\chi_2, \quad
  \psi_1\phi_1\varphi_2\chi_2+\psi_2\phi_2\varphi_1\chi_1, \nonumber \\
  &&& \psi_1\phi_2\varphi_1\chi_2+\psi_2\phi_1\varphi_2\chi_1, \quad
  \psi_1\phi_2\varphi_2\chi_1+\psi_2\phi_1\varphi_1\chi_2.
\end{alignat}
Their hermitian conjugates are also in the $1_S$ representations.

%%%%%%%%%%%%%%%%%%%%%%%%%%%%%%%%%%%%%%%%%%%%%%%%%%%%%%%%%%%%%%%%%%%%%%
\subsection{Group basis dependence}
\label{groupbasis}

We have presented the $S_3$ algebra in the complex basis where the
representation matrices are given by those in
Table~\ref{matrix}. There are, however, several bases of 
the $S_3$ matrices which are often used in the literature. So it may
be instructive here to describe the relation among these group bases,
compared to the complex basis used in the previous subsections.

%%%%%%%%%%%%%%%%%%%%%%%%%%%%%%%%%%%%%%%%%%%%%%%%%%%%%%%%%%%%%%%%%%%%%%
\subsubsection{Democratic basis}
\label{demobasis}

The democratic basis is adopted to produce a flavor-democratic mass
matrix in which all matrix elements are equal~\cite{S3demo}. In this
basis, the $S_3$ operations generate the permutations of three
objects, for example, the exchange of the first and second
indices. The invariance under such transformations require the
universal size of couplings for three generations if they belong to a
three-dimensional representation of $S_3$ and the Higgs field is in
the singlet.

Unlike in the complex basis, the representation matrices have
apparently non-trivial (not block-diagonal) structure for
three-dimensional representations. Different group bases are converted
to each other by unitary transformations. It is worth noting that
there are two types of democratic basis which correspond to the
existence of two three-dimensional 
representations; $3_{\rm S}$ and $3_{\rm A}$. The elements in the
democratic basis for $3_{\rm S}$ are defined by the unitary 
matrix $V$ as
\begin{equation}
  T^S_i \,=\, V T_i V^\dagger,
  \label{TS}
\end{equation}
where $T_i$'s in the right-handed side are given in the complex basis,
and $V\equiv UP_S$ with
\begin{equation}
  U \,=\, \begin{pmatrix}
    1/\sqrt{2} & 1/\sqrt{6} & 1/\sqrt{3} \\
    -1/\sqrt{2} & 1/\sqrt{6} & 1/\sqrt{3} \\
    0 & -2/\sqrt{6} & 1/\sqrt{3}
  \end{pmatrix}, \qquad
  P_S \,=\, \begin{pmatrix}
    1/\sqrt{2} & -1/\sqrt{2} & \\
    i/\sqrt{2} & i/\sqrt{2} & \\ & & \!1
  \end{pmatrix}.
  \label{U}
\end{equation}
Thus the representation matrices for $3_{\rm S}$ are given by the
label-changing matrices:
\begin{alignat}{6}
T^S_1 &= \begin{pmatrix}
  1 & & \\ & 1 & \\ & & 1
\end{pmatrix},& \qquad
T^S_2 &= \begin{pmatrix}
  & 1 & \\ & & 1 \\ 1 & &
\end{pmatrix},& \qquad
T^S_3 &= \begin{pmatrix}
  & & 1 \\ 1 & & \\ & 1 &
\end{pmatrix}, \nonumber \\
T^S_4 &= \begin{pmatrix}
  & 1 & \\ 1 & & \\ & & 1
\end{pmatrix},& \qquad
T^S_5 &= \begin{pmatrix}
  & & 1 \\ & 1 & \\ 1 & &
\end{pmatrix},& \qquad
T^S_6 &= \begin{pmatrix}
  1 & & \\ & & 1 \\ & 1 &
\end{pmatrix}.
\label{TS3S}
\end{alignat}
For the pseudo-triplet representation $3_{\rm A}$, the 
matrices $T^S_{1,2,3}$ have the same forms as 
above, but $T^S_{4,5,6}$ become rather complicated, as can be seen by
the definition (\ref{TS}).

Another democratic basis is defined so that the representation
matrices for $3_{\rm A}$ are expressed by label-exchanging matrices
like (\ref{TS3S}). The matrices $T^A_i$ are given by a similar unitary
rotation to (\ref{TS}), except that the matrix $P_S$ is now replaced by
\begin{equation}
  P_A \,=\, \begin{pmatrix}
    1/\sqrt{2} & 1/\sqrt{2} & \\
    i/\sqrt{2} & -i/\sqrt{2} & \\ & & \!1
  \end{pmatrix}.
\end{equation}
One can see that the matrices $T^A_{1,2,3}$ take the same forms as
those in (\ref{TS3S}) both for triplet and pseudo-triplet
representations. The differences appear 
for $T_{4,5,6}$; $T^A_{4,5,6}(3_{\rm A})=
-T^S_{4,5,6}(3_{\rm S})$ and $T^A_{4,5,6}(3_{\rm S})=
-T^S_{4,5,6}(3_{\rm A})$, the former matrices exchanges the three
objects as (\ref{TS3S}) and the latter have some complicated forms.

%%%%%%%%%%%%%%%%%%%%%%%%%%%%%%%%%%%%%%%%%%%%%%%%%%%%%%%%%%%%%%%%%%%%%%
\subsubsection{Real basis}

In the discussion of the democratic basis, $P_{S,A}$ rotate only the
first and second indices. This means that the representation 
matrices $T^{S,A}_i$ are rotated to block-diagonal forms only by 
the $U$ rotation, namely, the $U$-rotated matrices are decomposed into
the matrices for irreducible representations. (Note that $U$ is the
unitary rotation which diagonalizes the so-called flavor-democratic
mass matrix.) The real basis $T^U_i$ is defined by 
rotating $T^{S,A}_i$ with the $U$ matrix, for example,
\begin{equation}
  T^U_i \,=\, U^\dagger T^S_iU.
  \label{TU}
\end{equation}
These group elements take the following forms for 
the $3_{\rm S}$ representation:
\begin{alignat}{6}
T^U_1 &= \begin{pmatrix}
  1 & & \\ & 1 & \\\ & & 1
\end{pmatrix},& \quad
T^U_2 &= \begin{pmatrix}
  -1/2 & \sqrt{3}/2 & \\ -\sqrt{3}/2 & -1/2 & \\ & & 1
\end{pmatrix},& \quad
T^U_3 &= \begin{pmatrix}
  -1/2 & -\sqrt{3}/2 & \\ \sqrt{3}/2 & -1/2 & \\ & & 1
\end{pmatrix}, \nonumber\\
T^U_4 &= \begin{pmatrix}
  -1 & & \\ & 1 & \\ & & 1
\end{pmatrix},& \quad
T^U_5 &= \begin{pmatrix}
  1/2 & -\sqrt{3}/2 & \\ -\sqrt{3}/2 & -1/2 & \\ & & 1
\end{pmatrix},& \quad
T^U_6 &= \begin{pmatrix}
  1/2 & \sqrt{3}/2 & \\ \sqrt{3}/2 & -1/2 & \\ & & 1
\end{pmatrix},
\label{Treal}
\end{alignat}
which are block diagonal. It is clear from the definition that 
the $T^U$ basis is obtained from the complex basis by 
the $P_S$ or $P_A$ phase rotation. Note that, unlike in the complex
basis, all the representation matrices are real-valued. This is why we
call it the real basis. Due to this reality of matrix elements 
of $T^U_i$, two types of tensor products are possible 
for $S_3$ doublets. In the real basis of (\ref{TU}), the one tensor
product of two doublets is
\begin{eqnarray}
  \psi \times \phi &=& (\psi^\dagger\sigma_1\phi,\,
  \psi^\dagger\sigma_3\phi)_2^{\rm t}\,
  +(\psi^\dagger i\sigma_2\phi)_{1_{\rm A}}
  +(\psi^\dagger\phi)_{1_{\rm S}}, \\
  &=& (\psi_1^\dagger\phi_2+\psi_2^\dagger\phi_1,\,
  \psi_1^\dagger\phi_1-\psi_2^\dagger\phi_2)_2^{\rm t}\,
  +(\psi_1^\dagger\phi_2-\psi_2^\dagger\phi_1)_{1_{\rm A}}
  +(\psi_1^\dagger\phi_1+\psi_2^\dagger\phi_2)_{1_{\rm S}}.\quad
\end{eqnarray}
Another consistent product can be defined by using transposition 
of $\psi$, instead of $\psi^\dagger$. This choice is possible since
all the group elements are expressed in terms of real numbers. In the
language of the complex basis, the product defined with daggers
corresponds to (\ref{22}) and that with transpositions 
to (\ref{22c}). The difference between these two types of products
becomes important in the case that $S_3$-doublet fields transform
non-trivially under other symmetries than $S_3$. In particular, this
is indeed the case for the standard-model fermions and Higgs
bosons. The real basis has often been used in the
literature~\cite{S3real}. For completeness, we comment on another real
basis which is defined from $T^A_i$, instead 
of $T^S_i$ in (\ref{TU}). The matrices $T^U_i$ are now given 
by $T^U_i=U^\dagger T^A_iU$ and their explicit forms 
for $3_{\rm A}$ are (\ref{Treal}) by changing the signs 
of $T^U_{4,5,6}$ (i.e., $T^U_{4,5,6}\to -T^U_{4,5,6}$). Therefore the
non-trivial tensor product 
becomes $\psi\times\phi=(\psi^\dagger\sigma_3\phi,
-\psi^\dagger\sigma_1\phi)_2^{\rm t}
+(\psi^\dagger i\sigma_2\phi)_{1_{\rm A}}
+(\psi^\dagger\phi)_{1_{\rm S}}$ or that with $\psi^{\rm t}$.

Finally we show the singlet components contained in the products of
more than two doublets. In the real basis, they are given by (up to
the fourth order)
\begin{alignat}{3}
  &\psi\times\phi &\;\;\;\supset\;\;\;
  &\psi_1\phi_1+\psi_2\phi_2, \\
  &\psi\times\phi\times\varphi &\supset\;\;\;
  &\psi_1\phi_1\varphi_2+\psi_1\phi_2\varphi_1
  +\psi_2\phi_1\varphi_1-\psi_2\phi_2\varphi_2, \\
  &\psi\times\phi\times\varphi\times\chi &\supset\;\;\;
  &(\psi_1\phi_1+\psi_2\phi_2)(\varphi_1\chi_1+\varphi_2\chi_2), \quad
  (\psi_1\varphi_1+\psi_2\varphi_2)(\phi_1\chi_1+\phi_2\chi_2),
  \nonumber \\
  &&&(\psi_1\chi_1+\psi_2\chi_2)(\phi_1\varphi_1
  +\phi_2\varphi_2), \quad (\psi_1\phi_2-\psi_2\phi_1)
  (\varphi_1\chi_2-\varphi_2\chi_1), \nonumber \\
  &&&(\psi_1\varphi_2-\psi_2\varphi_1)(\phi_1\chi_2
  -\phi_2\chi_1), \quad (\psi_1\chi_2-\psi_2\chi_1)
  (\phi_1\varphi_2-\phi_2\varphi_1).
\end{alignat}
In the right-handed sides, the 
components $\psi_i$, $\phi_j$, $\cdots$ can be replaced 
with $\phi_i^\dagger$, $\phi_j^\dagger$, $\cdots$. Namely, 
$(\psi_1^\dagger\phi_1+\psi_2^\dagger\phi_2)$,
$(\psi_1\phi_1^\dagger\varphi_2+\psi_1\phi_2^\dagger\varphi_1
+\psi_2\phi_1^\dagger\varphi_1-\psi_2\phi_2^\dagger\varphi_2)$, and
others are also $S_3$ singlets. This is due to the fact that the
representation matrices $T^U_i$ are real in the real basis.

%%%%%%%%%%%%%%%%%%%%%%%%%%%%%%%%%%%%%%%%%%%%%%%%%%%%%%%%%%%%%%%%%%%%%%
\section{Invariant matrices}
\label{inv}

Various forms of mass textures of quarks and leptons have been known
to be phenomenologically viable~\cite{Fritzsch}-\cite{FGM}. As
discussed in the previous section, $S_3$ has three types of
irreducible representations; $2$, $1_{\rm A}$ and $1_{\rm S}$. We
would like here to examine what combinations of $S_3$ representations
for matter fields lead to those mass textures. Higgs fields may also
transform non-trivially under the $S_3$ symmetry and play significant
roles for realizing textures forms. Since there are three repetitions
of matter fields, various assignments of $S_3$ representations are
available, where the first two generations constitute a doublet and so
on. These include a possibility to realize asymmetrical forms of mass
textures and give rise to physically observable effects such as
flavor-violating processes in future experiments. In particular, as
mentioned in the introduction, vanishing matrix elements lead to
interesting phenomenological consequences. Such zero elements could be
obtained in the framework of non-abelian discrete flavor symmetry by
suitably taking Higgs representations and their VEV forms, which are
calculable by analyzing scalar potential. Natural forms of VEVs
generally depend on the $S_3$ group basis in which mass matrices are
described. At this stage, explicit forms of mass textures are governed
only by discrete flavor symmetry. In some cases, however, additional
implementation might be needed to have more control, in particular,
for fermion mass hierarchy being naturally realized. In the present
framework, the key ingredients for model construction are to 
select (i) $S_3$ group structure, (ii) representations of matter 
fields, (iii) Higgs profiles (representations 
and VEVs), and (iv) extra symmetries. Let us first comment on these
issues in some details.

\smallskip

(i) $S_3$ group structures:~ 
As mentioned in Section~\ref{groupbasis}, physical consequences of a
mass texture potentially depend on the group basis adopted in 
constructing $S_3$ models. It is noted that a choice of specific basis
does not affect physical results as long as the flavor symmetry is
unbroken: apparent basis dependence, e.g.\ different forms of mass
matrices, is only due to a choice of flavor 
basis. The $S_3$ invariance guarantees the same spectrum without
regard to basis choices. However the predictions for generation mixing
might be different. This is because, in the standard model, there
already exists a basis which defines the generation structure, namely
the interaction basis where the weak current interaction is flavor
diagonal. Once one picks up a group basis from some model-building
perspective, the relabeling of flavor indices by $S_3$ transformation
gives physical meaning on generation structure. Another important
basis dependence appears in breaking the flavor symmetry. Since any
kind of flavor symmetry has not been observed in the low-energy
regime, in principle any form of symmetry-breaking patterns is
possible. In realistic model construction, some guiding principles are
often adopted, such as simplicity and/or dynamical justification. In
either case, symmetry-breaking parameters depend on 
the $S_3$ basis. While the breaking parameters take a simple form in
one basis, they are rotated to a complicated form in other general
bases, which form seems to be completely unnatural. In this way, the
choice of flavor-group basis may have physical consequences if
three-generation fermions are assigned to (pseudo) triplets, and
therefore is an important factor in constructing models with flavor
symmetry.

Another issue is what types of $S_3$ symmetries are involved into the
theory. For example, with only one $S_3$ symmetry, both left- and
right-handed fermions (and also Higgs bosons) transform under the 
same $S_3$. On the other hand, one may easily imagine that
three-generation fermions have non-trivial charges of 
separate $S_3$ groups. A well-known example of the latter case is the
democratic mass texture~\cite{S3demo} realized 
by $S_3{}_L\times S_3{}_R$ symmetries.

\smallskip

(ii) Representations of matter fields:~
In this paper, $S_3$ is introduced to control the flavor structure of
quarks and leptons. Three-generation matter fields generally belong to
non-trivial representations of such flavor symmetries. A charge
assignment often assumed in the literature is that the first and
second generations make up a doublet. This assignment is adopted to
account for several phenomenological issues. First, the flavor
symmetry invariance suppresses flavor-violating effects between the
first and second generations, which effects have been tightly
constrained by various experimental results in supersymmetric
theory~\cite{FCNC}. Second, if the third generation has different
flavor charges from the other twos, they are appropriate to have
larger masses. It is however noted that all other flavor charge
assignments are equally allowed. For example, a phenomenologically
interesting alternative is that the second and third generation
leptons make an $S_3$ doublet. This is motivated by the recent
observations of neutrino flavor mixing between the second and third
generations. Moreover it could account for the lightness of the
first-generation fermions. In what follows, we show that
representations of matter fields, including left-right asymmetric
assignments, are useful for obtaining various types of mass matrices.

\smallskip

(iii) Higgs profiles:~ Phenomenologically indirect but important
possibilities arise for the profile of Higgs 
fields. If $SU(2)_W$-doublet Higgses are in non-trivial
representations of $S_3$, Yukawa couplings can be described by
renormalizable operators. Higher-dimensional operators including
multiple Higgs fields are suppressed by a large cutoff scale and give
negligible corrections to Yukawa couplings, which is a nice feature in
a sense that everything is described within the renormalizable
level. This approach however requires that hierarchically small values
of Yukawa couplings for the first two generations must be fixed by
hand as in the standard model. Moreover, due to the existence of
multiple $SU(2)_W$-doublet Higgses, naive gauge coupling unification
is spoiled, and Higgs-mediated flavor-violating effects might not be
negligible even at tree level. An alternative choice of Higgs charges
is that $SU(2)_W$-doublet Higgses belong to the flavor singlet. In
this case, Yukawa interactions are effectively derived from
higher-dimensional operators which can be made invariant by
introducing appropriate scalars $\Phi$'s in 
non-trivial $S_3$ representations. The VEVs of $\Phi$ break the flavor
symmetry and generate trilinear Yukawa couplings below the breaking
scale. Since the ratio $\langle\Phi\rangle$ to a cutoff scale gives a
unit of Yukawa hierarchy of quarks and leptons, $S_3$ should be broken
at a high-energy scale below the cutoff. This situation resolves the
above-mentioned problems of $S_3$-charged $SU(2)_W$ Higgs 
doublets: Yukawa hierarchy is explained by controlled
higher-dimensional operators, gauge coupling unification is 
preserved, and Higgs-mediated flavor violation is suppressed by a
large $S_3$-breaking scale. As an imprint of such high-scale flavor
symmetry, new sources of flavor violation could arise from
renormalization-group evolution below the symmetry-breaking scale. For
example, if the theory is supersymmetrized, flavor-changing couplings
are generally induced, and their magnitudes depend on the dynamics of
supersymmetry breaking. While the flavor violation tends to be small
as K\"ahler terms are limited by flavor symmetry, it may be
observable, e.g.\ in the gravity mediation
scenario~\cite{gravity}. Note that this type of flavor violation is
negligible in the case of $S_3$-charged Higgs bosons as long as the
flavor symmetry remains intact at low-energy regime.

The $S_3$ property of Higgs fields is also relevant to the group
basis. That is the form of VEVs and their naturalness in the sense 
of 't~Hooft. For example, if a pseudo-singlet field develops a
non-vanishing VEV, $S_3$ is broken down to a subgroup $Z_3$. In the
limit of other VEVs being zeros, low-energy effective theory still has
the residual $Z_3$ invariance. In fact, for realistic 
cases, some $S_3$ doublets have nonzero VEVs in order to give
non-trivial flavor structure. Then the maximal residual subgroup 
is $S_2$. Notice here that the $S_2$-invariant forms of VEVs depend on
the group basis. It is found from (\ref{Tcpx}) and (\ref{Treal}) that
such technically natural VEV of $S_3$ doublet is proportional 
to $(1,1)^{\rm t}$ in the complex basis 
and $(0,1)^{\rm t}$ [or $(1,0)^{\rm t}$] in the real basis. The former
is available to realize large flavor mixing and the latter is useful
for generating textures with vanishing elements. Since the flavor
symmetry is completely broken at low energy, the group basis might be
chosen so that symmetry-breaking forms seem as natural as
possible. Finally, there is another form of doublet VEV which is often
utilized in the literature. That 
is $(x,1)^{\rm t}$ with $x\ll 1$, which arises from a linear
combination of the above two types of VEVs, but it is nothing but
parameter tuning.

\smallskip

(iv) Extra symmetries:~
The $S_3$ flavor symmetry does not fully explain the mass hierarchies
of quarks and leptons. A hierarchical order of $S_3$-breaking VEVs
requires different order of couplings whose origin is generally
unclear. An attractive way to dynamically justify hierarchical
couplings is to introduce extra symmetries into the theory. As for
Yukawa couplings, a well-known example is the Froggatt-Nielsen
mechanism~\cite{FN} where extra symmetry controls the orders of Yukawa
couplings. They are generated in low-energy effective theory via
decoupling heavy fields, i.e.\ higher-dimensional operators whose
coefficients are naturally given by the fundamental scale of the theory.

\smallskip

In the following, we present several examples of mass matrices which
stem from the $S_3$ flavor symmetry. In almost the cases,
three-generation fermions belong to non-trivial triplet
representations: two of them make up an $S_3$ doublet and the other is
a (pseudo) singlet. On the other hand, appropriate representations of
Higgs fields are chosen to have non-vanishing elements 
in $S_3$-invariant Yukawa matrices. Mass textures will be expressed in
the complex basis unless particularly mentioned. The expressions in
the other bases are easily obtained by the unitary basis rotations
defined in the previous section.

%%%%%%%%%%%%%%%%%%%%%%%%%%%%%%%%%%%%%%%%%%%%%%%%%%%%%%%%%%%%%%%%%%%%%%
\subsection{A single $\boldsymbol{S_3}$}

First we study the case that both left-handed 
fermions $\psi_L{}_i$ and right-handed 
ones $\psi_R{}_j$ ($i,j=1,2,3$) transform under a 
single $S_3$ group. This case is also straightforwardly applied to
Majorana mass terms with the identification $\psi_L=\psi_R$. In the
following, we take a charge assignment 
that $(\psi_L{}_1,\psi_L{}_2)$ and $(\psi_R{}_1,\psi_R{}_2)$ 
are $S_3$ doublets and the other fermions are in (pseudo) singlet
representation. This does not lose any generalities since mass
matrices for other charge assignments are obtained by flavor
rotations. As for Higgs fields, all the three types of irreducible
representations are available for giving nonzero matrix 
elements; a doublet $H_D=(H_1,H_2)$, a pseudo 
singlet $H_A$, and a singlet $H_S$. Here $H_{D,S,A}$ are interpreted
as either $SU(2)_W$-doublet elementary scalars or products of some
numbers of fields [see, e.g.\ the comment (iii) Higgs profiles in the
above].

%%%%%%%%%%%%%%%%%%%%%%%%%%%%%%%%%%%%%%%%%%%%%%%%%%%%%%%%%%%%%%%%%%%%%%
\subsubsection{The general case}
\label{generalS3}

Dirac mass terms flip the chirality of fermions. Given 
that $\psi_L{}_3$ and $\psi_R{}_3$ are singlets, the most 
general $S_3$-invariant Dirac mass matrix is described as
\begin{gather}
  {\cal L}_{\rm Dirac} \,=\, -\overline{\psi_R{}_i}\,M_{ij}
  \,\psi_L{}_j +{\rm h.c.}, \\[1mm]
  M \,=\, \begin{pmatrix}
  aH_S+a'H_S^*+bH_A+b'H_A^* & cH_2+c'H_1^* & dH_1+e'H_2^*\, \\
  cH_1+c'H_2^* & aH_S+a'H_S^*-bH_A-b'H_A^* & dH_2+e'H_1^*\, \\ 
  eH_2+d'H_1^* & eH_1+d'H_2^* & fH_S+f'H_S^*\,
  \end{pmatrix},
  \label{MD}
\end{gather}
where $a,a',b,\cdots,f,f'$ are independent coupling constants. The
generic form of mass matrix (\ref{MD}) is simplified 
if $H_{D,S,A}$ belong to complex representations of other groups 
than $S_3$: either $H_x$ or corresponding $H_x^*$ is dropped out in
each element of the generic matrix. This is indeed the case for 
the $SU(2)_W$ Higgs doublets in the standard model. Another charge
assignment is that the pseudo singlet representation $1_{\rm A}$ is
adopted for the third-generation fermions. For 
example, when $\psi_L{}_3$ is a pseudo singlet, the third column 
of (\ref{MD}) is modified so that $H_2$ and $H_S$ are replaced 
with $-H_2$ and $H_A$, respectively.

The general form of Majorana mass matrix is described by 
identifying $\psi_L=\psi_R\equiv\psi$. It is noted that the tensor
product (\ref{22c}) should be used in constructing Majorana mass term
in the complex basis. Thus the $S_3$ invariance leads to
\begin{gather}
  {\cal L}_{\rm Majorana} \,=\, -\frac{1}{2}\,
  \overline{\psi^c_{\,i}}\,M_{ij}\,\psi{}_j +{\rm h.c.}, \\[1mm]
  M \,=\, \begin{pmatrix}
  \,aH_1+bH_2^* & cH_S+c'H_S^* & dH_2+eH_1^*\, \\
  \,cH_S+c'H_S^* & aH_2+bH_1^* & dH_1+eH_2^*\, \\
  \,dH_2+eH_1^* & dH_1+eH_2^* & fH_S+f'H_S^*\,
  \end{pmatrix}.
  \label{MR}
\end{gather}
Similar to the above case of Dirac masses, either $H_x$ or $H_x^*$ is
removed in each matrix element when $H_x$ has some (complex) quantum
charge other than that of $S_3$. Due to the difference of $S_3$ tensor
products used for Dirac and Majorana mass terms, the resulting flavor
structures in (\ref{MD}) and (\ref{MR}) are rather different. This
fact could provide an interesting possibility for realistic model
construction of flavor.

%%%%%%%%%%%%%%%%%%%%%%%%%%%%%%%%%%%%%%%%%%%%%%%%%%%%%%%%%%%%%%%%%%%%%%
\subsubsection{Supersymmetric case}

Supersymmetry invariant fermion masses come from the 
superpotential, which is a holomorphic function of chiral
superfields. That is, superpotential terms are written in terms of
chiral superfields, which contain fermionic components with a certain
chirality, e.g.\ left-handed fermions. In this case, right-handed
fermions are introduced as charge conjugations of left-handed ones. As
in the general case, suppose that the chiral superfields of first two
generations, $(\Psi_L{}_1,\Psi_L{}_2)$ 
and $(\Psi_R{}_1,\Psi_R{}_2)$, are $S_3$ doublets, 
and $\Psi_R{}_3$, $\Psi_L{}_3$ are singlets. It should be noted that
this assignment leads to the right-handed 
fermions $(\psi_R{}_1,\psi_R{}_2)$ being an anti-doublet. Therefore
the product (\ref{22c}) is applied to constructing $S_3$-invariant
superpotential. The most general supersymmetric mass term is thus
given by
\begin{gather}
  W_{\rm Dirac} \,=\, \Psi_R{}_i\,M_{ij}\,\Psi_L{}_j, \\[1mm]
  M \,=\, \begin{pmatrix}
  aH_1 & bH_S+cH_A & dH_2 \\
  bH_S-cH_A & aH_2 & dH_1 \\
  eH_2 & eH_1 & fH_S
  \end{pmatrix}.
  \label{MDSUSY}
\end{gather}
In the case that $\Psi_L{}_3$ is assigned to a pseudo singlet, one
needs the replacement $H_2\to-H_2$ and $H_S\to H_A$ in the third
column of (\ref{MDSUSY}). Similarly, the $S_3$-invariant Majorana mass
term is easily found by symmetrizing the 
matrix (\ref{MDSUSY}), namely, by setting the couplings 
as $c=0$ and $d=e$. This does not significantly modify the flavor
structure, unlike in the non-supersymmetric case.

%%%%%%%%%%%%%%%%%%%%%%%%%%%%%%%%%%%%%%%%%%%%%%%%%%%%%%%%%%%%%%%%%%%%%%
\subsubsection{Relation to the democratic mass matrix}

We have descried the mass matrices in the complex basis of 
the $S_3$ group. Let us here comment on the relation to the so-called
democratic form of mass matrix in which all the matrix elements have
equal magnitude. It has a simple $S_3$ derivation in the democratic
basis defined in Section~\ref{demobasis}. As we explained 
before, these two group bases are connected by the unitary 
rotation $U$ in (\ref{U}). A simple way to recover the democratic mass
matrix is to introduce only an $S_3$-singlet Higgs field $H_S$. As
seen from (\ref{MD}), that corresponds to a flavor-diagonal matrix in
the complex basis. By rotating the matrix to the democratic basis, it
turns out to be
\begin{equation}
  M \,=\, \begin{pmatrix}
  1 & 1 & 1 \\ 1 & 1 & 1 \\ 1 & 1 & 1
  \end{pmatrix} h_S + \begin{pmatrix}
    1 & & \\ & 1 & \\ & & 1 
  \end{pmatrix} h_S',
\end{equation}
where $h_S=[(f-a)H_S+(f'-a')H_S^*]/3$ 
and $h_S'=aH_S+a'H_S^*$. It is not hard to see the $S_3$ invariance of
the above two matrices in the democratic basis. The democratic mass
matrix is thus found to be derived from an assumption that only one
matrix element is dominant in the complex basis. Therefore a 
single $S_3$ symmetry cannot ensure the flavor democracy: even with a
single $H_S$ field, the democratic ansatz is generally disturbed. As
seen from the representation matrices (\ref{TS3S}), the democratic
basis deals with the three indices of a (pseudo) triplet equivalently,
and is not suitable to discuss the decomposition to irreducible
representations.

%%%%%%%%%%%%%%%%%%%%%%%%%%%%%%%%%%%%%%%%%%%%%%%%%%%%%%%%%%%%%%%%%%%%%%
\subsection{$\boldsymbol{S_3{}_L\times S_3{}_R}$}

For controlling chirality-flipping operators, one can utilize a flavor
symmetry under which left-handed and right-handed fermions transform
separately, that is, the $S_3{}_L\times S_3{}_R$ symmetry. Without
loss of any generalities, we assume that the first two 
generations $(\psi_L{}_1,\psi_L{}_2)$ 
and $(\psi_R{}_1,\psi_R{}_2)$ belong to the doublet representations 
of $S_3{}_L$ and $S_3{}_R$, respectively. Since there exist three
irreducible representations for the $S_3$ group, fermion bilinear
terms transform in nine ways under $S_3{}_L\times S_3{}_R$. The
corresponding nine types of scalars which ensures the flavor
invariance are denoted by $H_{ij}$ ($i,j=D,S,A$), where an obvious
notation has been used, e.g.\ $H_{DA}$ means a doublet 
under $S_3{}_L$ and a pseudo singlet of $S_3{}_R$. As mentioned
before, the symbols $H_{ij}$ stand for either $SU(2)_W$-doublet
elementary scalars or products of some fields with appropriate charges.

%%%%%%%%%%%%%%%%%%%%%%%%%%%%%%%%%%%%%%%%%%%%%%%%%%%%%%%%%%%%%%%%%%%%%%
\subsubsection{The general case}

For an illustration, we assume that the third-generation 
fermions $\psi_L{}_3$ and $\psi_R{}_3$ are singlets 
of $S_3{}_L$ and $S_3{}_R$, respectively. In case that they are pseudo
singlets, the subscripts $S$ of Higgs fields should be replaced 
with $A$ in the following expressions. The most general Dirac mass
matrix is made symmetry invariant by 
introducing $H_{DD}$, $H_{DS}$, $H_{SD}$, and $H_{SS}$, and is given by
\begin{gather}
  {\cal L}_{\rm Dirac} \,=\, -\overline{\psi_R{}_i}\,M_{ij}
  \,\psi_L{}_j +{\rm h.c.}, \\[2mm]
  M \,=\, \begin{pmatrix}
   \,\begin{minipage}{4.2cm}
   $a(H_{DD})_{11}+b(H_{DD})_{12} \\[1mm]
   ~ +c(H_{DD}^{\,*})_{21}+d(H_{DD}^{\,*})_{22}$ 
 \end{minipage} &
  \begin{minipage}{4.2cm} 
   $a(H_{DD})_{12}+b(H_{DD})_{11} \\[1mm]
   ~ +c(H_{DD}^{\,*})_{22}+d(H_{DD}^{\,*})_{21}$ 
  \end{minipage} & 
  e(H_{SD})_1 +f(H_{SD}^{\,*})_2 \\[5mm]
  \begin{minipage}{4.2cm} 
   $a(H_{DD})_{21}+b(H_{DD})_{22} \\[1mm]
   ~ +c(H_{DD}^{\,*})_{11}+d(H_{DD}^{\,*})_{12}$
  \end{minipage} & 
  \begin{minipage}{4.2cm}
   $a(H_{DD})_{22}+b(H_{DD})_{21} \\[1mm]
   ~ +c(H_{DD}^{\,*})_{12}+d(H_{DD}^{\,*})_{11}$
  \end{minipage} & 
  e(H_{SD})_2 +f(H_{SD}^{\,*})_1 \\[5mm]
  g(H_{DS})_2 +h(H_{DS}^{\,*})_1 & g(H_{DS})_1 +h(H_{DS}^{\,*})_2 & 
  jH_{SS}+j'H_{SS}^{\,*} 
  \end{pmatrix},
  \label{generalMD}
\end{gather}
where $a,b,\cdots,j,j'$ are the coupling constants. The Higgs fields
either with or without asterisks are dropped out in each matrix
element if they have some quantum numbers of other symmetries 
than $S_3{}_L\times S_3{}_R$. The possible Majorana mass term 
of $\psi_L$ (or $\psi_R$) is written in the same way as (\ref{MR}) by
including other $H$'s with appropriate charges.

The Dirac mass term in supersymmetric theory is given by the
superpotential which is a analytic function of superfields with
definite chirality. If we take the same flavor charge assignment as
above for the three-generation 
superfields $\Psi_L{}_i$ and $\Psi_R{}_j$, the most general form of
mass matrix is
\begin{gather}
  W_{\rm Dirac} \,=\, \Psi_R{}_i\,M_{ij}\,\Psi_L{}_j, \\[1mm]
  M \,=\, \begin{pmatrix}
  a(H_{DD})_{22} & a(H_{DD})_{21} & b(H_{SD})_2 \\
  a(H_{DD})_{12} & a(H_{DD})_{11} & b(H_{SD})_1 \\
  c(H_{DS})_2 & c(H_{DS})_1 & dH_{SS}
  \end{pmatrix}.
\end{gather}

%%%%%%%%%%%%%%%%%%%%%%%%%%%%%%%%%%%%%%%%%%%%%%%%%%%%%%%%%%%%%%%%%%%%%%
\subsubsection{Examples}

It is found from (\ref{generalMD}) that the general form of Dirac mass
term is rather complicated. Combining with simple assumptions, we here
present several examples where phenomenologically interesting forms of
mass matrices are obtained, in particular, by choosing relevant Higgs
contents.

The first example is the introduction of a single elementary 
scalar $H$ in the (pseudo) singlet representation of
both $S_3{}_L$ and $S_3{}_R$ symmetries. According to 
the $S_3$ charges of $\psi_{L3,R3}$, a relevant representation of the
scalar becomes that of $H_{ij}$ with $i,j=$ $S$ or $A$. In any case,
only the 3-3 element of mass matrix is allowed;
\begin{equation}
  M \,=\, \begin{pmatrix}
   ~ & ~ & \\ & & \\ & & aH
  \end{pmatrix}.
  \label{Mcpx}
\end{equation}
If turning to the democratic basis, we have
\begin{equation}
  M \,=\, \frac{1}{3}\begin{pmatrix}
    1 & 1 & 1 \\ 1 & 1 & 1 \\ 1 & 1 & 1
  \end{pmatrix} aH.
  \label{Mdemo}
\end{equation}
Both of these matrices have only one nonzero eigenvalue and the same
mass spectrum. However the resulting flavor mixing structures are
clearly different: the matrix (\ref{Mdemo}) induces large flavor
mixing, while (\ref{Mcpx}) has no mixing between the third and the
first-two generations. As we noted before, this may be a physical
difference because, in realistic models, there exists more than one
sector whose relative basis differences do affect flavor structure
such as the quark mixing angles.

The second example is not to consider that elementary 
scalars $H_{ij}$ give all trilinear Yukawa couplings, but to work with
some products of scalars in lower-dimensional representations. As an
example, let us introduce three scalar 
fields $H_D=(H_1,H_2)$, $H_D'=(H_1',H_2')$ and $H_S$ whose
representations of ($S_3{}_L, S_3{}_R$) 
are ($2,1_{\rm S}$), ($1_{\rm S},2$), and ($1_{\rm S},1_{\rm S}$),
respectively. With this field content and the matter representation as
above, all entries in the Dirac mass matrix can be filled up with
non-renormalizable operators. In supersymmetric theory, the mass
matrix is now given by
\begin{equation}
  M \,=\, \begin{pmatrix}
   aH_2H_2' & aH_1H_2' & bH_2' \\
   aH_2H_1' & aH_1H_1' & bH_1' \\
   cH_2 & cH_1 & dH_S
  \end{pmatrix}.
  \label{MHH}
\end{equation}
On top of economical field content, this example has several useful
properties for constructing flavor models of quarks and
leptons. First, the matrix (\ref{MHH}) has a vanishing determinant and
therefore provides a compelling dynamical reason for the observed tiny
masses of first-generation fermions. Secondly, the effective Yukawa
couplings for the first-two generations become naturally small. This
is because the $S_3$ invariance requires that they come from
higher-dimensional operators suppressed by some large mass
scale. Thirdly, the hierarchical flavor structure is easily attained
with a smaller number of scalar VEVs. It might often lead to some
relations among mass eigenvalues and mixing angles of quarks and
leptons. One can make use of these interesting features with a smaller
number of representations of scalar fields than (\ref{generalMD}).

%%%%%%%%%%%%%%%%%%%%%%%%%%%%%%%%%%%%%%%%%%%%%%%%%%%%%%%%%%%%%%%%%%%%%%
\subsection{Singlet flavor}

We have so far discussed the case that both left- and right-handed
fermions have non-trivial charges of discrete flavor symmetry. In this
subsection, we comment on a possibility that either left- or
right-handed generations is insensitive to flavor transformation.

Let us consider a single $S_3$ symmetry under which left-handed
generations transform non-trivially but right-handed ones do not. In a
similar way to the discussion in Section~\ref{generalS3}, the generic
types of representations, $H_D=(H_1,H_2)$, $H_S$, and $H_A$, are taken
into account. If one takes a charge assignment 
that $(\psi_L{}_1,\psi_L{}_2)$ is a doublet, the most general Dirac
mass matrix is given by
\begin{gather}
  {\cal L}_{\rm Dirac} \,=\, 
  -\overline{\psi_R{}_i}\,M_{ij}\,\psi_L{}_j +{\rm h.c.}, \\[1mm]
  M \,=\, \begin{pmatrix}
  \,aH_2+bH_1^* & aH_1+bH_2^* & gH_S+g'H_S^*\, \\
  \,cH_2+dH_1^* & cH_1+dH_2^* & hH_S+h'H_S^*\, \\
  \,eH_2+fH_1^* & eH_1+fH_2^* & jH_S+j'H_S^*\, 
  \end{pmatrix}.
\end{gather}
In each matrix element, a symbol $H$ either with or without asterisk
should be dropped if $H$ is accompanied with some charges other than
that of $S_3$. In supersymmetric theory, the mass matrix is simplified to
\begin{equation}
  M \,=\, \begin{pmatrix}
  aH_2 & aH_1 & gH_S \\
  cH_2 & cH_1 & hH_S \\
  eH_2 & eH_1 & jH_S
  \end{pmatrix}.
  \label{3s}
\end{equation}
If only a singlet Higgs field exists, that results in
\begin{equation}
(\ref{3s}) \,\to\, \begin{pmatrix}
  g & g & g \\
  h & h & h \\
  j & j & j 
\end{pmatrix}H_S
\end{equation}
in the democratic basis. This form of Dirac mass matrix has been
discussed in the so-called lopsided models. Similar results are also
obtained with the $1_{\rm A}$ representation instead of $1_{\rm S}$.

%%%%%%%%%%%%%%%%%%%%%%%%%%%%%%%%%%%%%%%%%%%%%%%%%%%%%%%%%%%%%%%%%%%%%%
\section{Mass textures from $\boldsymbol{S_3}$}
\label{massS3}

On the prescription described in the previous sections, we will
perform several constructions of mass textures by use of 
flavor $S_3$ symmetry. That includes well-established forms of mass
matrices which have been discussed in the literature. In realistic
flavor models for quarks and leptons, some combinations of texture
forms are usually assumed, and it is therefore meaningful to examine
whether they can be simultaneously reproduced by horizontal symmetry.

It is found in Section~\ref{inv} that the key ingredients 
for $S_3$ model building are to define matter and Higgs profiles and,
if needed, to introduce additional symmetry. Among such further
symmetries, we will focus in this work on $U(1)$ flavor symmetry to
dynamically realize fermion mass hierarchy~\cite{FN,U1}. It 
is, however, highly non-trivial to assign relevant $U(1)$ quantum
numbers. This is because an $S_3$ doublet, e.g.\ the first and second
generations in the above matrix examples, must have the 
same $U(1)$ charge and it does not provide mass hierarchies between
them. Moreover, a naive charge assignment often leads to a prediction
that classical mass hierarchy realized by flavor symmetry is disturbed
by higher-dimensional operators. These facts are typical features 
of $S_3$ flavor models accompanied by additional $U(1)$. In models
without discrete flavor symmetry, $U(1)$ charge assignment generally
has wider flexibility~\cite{U1}. On the other hand, models with 
only $S_3$ symmetry have to do with fermion mass hierarchy by
arbitrarily tuning parameters such as Yukawa couplings and 
Higgs VEVs. In the following, we illustrate several examples of mass
textures derived from $S_3$ and assistant $U(1)$ flavor dynamics, and
also mention how to cure the above problems of charge assignment. It
is noticed that mass hierarchy is dynamically realized in a similar
way with a discrete subgroup of the flavor $U(1)$ symmetry. For 
example, all the hierarchical mass textures we discuss below can also
be obtained by $Z_N$ subgroup with appropriate (enough large) $N$ and
the same quantum numbers as in the $U(1)$ case. A smaller choice 
of $N$ would be possible and interesting from a viewpoint of
brevity. In this case, the problem of fermion masses can be handled
with flavor symmetries that are entirely discrete.

In what follows, we consider the cases that the flavor symmetries are
broken at some high-energy scale, and renormalization-group running
down to low energy should be taken into account if one obtains precise
values of coupling constants, once the model below the
symmetry-breaking scale is specified. However, it is not hard to see
that, in all examples we discuss below, the running effects do not
change qualitative results and can safely be dropped. The analysis in
this section is performed in the models with a single 
flavor $S_3$ symmetry and the results are described in the complex
basis, unless we particularly mention it.

%%%%%%%%%%%%%%%%%%%%%%%%%%%%%%%%%%%%%%%%%%%%%%%%%%%%%%%%%%%%%%%%%%%%%%
\subsection{Nearest neighbor form}

The first example of $S_3$ models includes the mass texture proposed
by Fritzsch~\cite{Fritzsch}. We assign non-trivial $S_3$ charges to
three-generation left- and right-handed 
fermions $\psi_L{}_i$ and $\psi_R{}_j$ ($i=1,2,3$). They are given 
by $(\psi_L{}_1,\psi_L{}_2)+\psi_L{}_3$ 
and $(\psi_R{}_1,\psi_R{}_2)+\psi_R{}_3$, namely, the first and
second generations make up $S_3$ doublets and the third ones are
singlets.

For the Higgs profile, we introduce 
an $S_3$-doublet $H_D=(H_1,H_2)$ as only a scalar with 
non-trivial $S_3$ charge. As mentioned in Section~\ref{inv}, $H_D$ has
two possibilities concerned with the electroweak charge. We here 
take $H_D$ as a singlet of the electroweak gauge symmetry, and
accordingly utilize the usual $SU(2)_W$-doublet Higgs $h$. Advantages
of this choice are the suppression of flavor-changing rare processes,
the preservation of gauge coupling unification in supersymmetric
theory, and so on. A key point is that the VEV of the $S_3$ doublet
scalar is assumed to take the 
form $\langle H_D\rangle=(\langle H_1\rangle,\,0)$. This form can be
dynamically justified by analyzing the scalar potential, as we will
show in later section. Another type of 
VEV, $\langle H_D\rangle=(0,\langle H_2\rangle)$, just gives the case
easily found by exchanging the up and down components in $S_3$
doublets.

The next step is to impose extra symmetries. As a simple example, we
here assume a $Z_2$ parity acting non-trivially 
on $S_3$ doublets (``doublet parity''). We also have 
flavor $U(1)$ symmetry in order to control mass hierarchy and mixing
among the three generations. Notice that due to the assignment of
these non-vanishing charges the resultant mass matrices are
simplified. That is, either a Higgs field or its complex conjugate can
appear in each matrix element. At this stage, the $3\times 3$ mass
matrix for $\psi_{L,R}$ is found to be
\begin{equation}
  \begin{pmatrix}
    & a & \\ a & & b\langle H_1\rangle\!\\ & c\langle H_1\rangle & d    
  \end{pmatrix}h,
  \label{NNI}
\end{equation}
where $a$, $b$, $c$, and $d$ are the ${\cal O}(1)$ coefficients which
include coupling constants and, if any, scalar fields with
non-vanishing VEVs. Throughout this paper, ${\cal O}(1)$ coefficients
are denoted where the fundamental mass scale of theory is taken to be
unity. The texture form (\ref{NNI}) was first adopted by
Fritzsch~\cite{Fritzsch} for quark mass matrices. It is noticed that
the parity leads to some of vanishing matrix elements at tree
level. An alternative parity assignment is viable 
where $\psi_L{}_3$, $\psi_R{}_3$ and $H_D$ have negative parity. A
more interesting possibility is to incorporate supersymmetry in the
model. In this case, vanishing matrix elements could be due to the
holomorphicity of superpotential, combined with $U(1)$ quantum charges
for realizing mass hierarchy, i.e.\ supersymmetric texture zeros.

To generate fermion mass hierarchy, we use the flavor $U(1)$ symmetry
whose quantum charges are assigned as in the table below:
\begin{equation}
  \begin{array}{c|ccccccc}
    & \psi_L{}_{1,2} & \psi_L{}_3 & \psi_R{}_{1,2} & \psi_R{}_3 & 
    h & H_D & \varphi \\ \hline
    U(1) & x & x' & y & y' & 0 & z & -1 \\[1mm]
    Z_2 & - & + & - & + & + & - & +
  \end{array}\nonumber
\end{equation}
We have introduced a scalar field $\varphi$ to describe
symmetry-invariant higher-dimensional operators which effectively
induce mass terms below the $U(1)$ breaking 
scale $\langle\varphi\rangle$. The $U(1)$ charges are normalized by
letting $\varphi$ take a unit charge, and $h$ is assumed to be neutral
in order to fix overall mass scale. As a result, the orders of
magnitude of the matrix elements in (\ref{NNI}) becomes
\begin{equation}
  \begin{pmatrix}
    & \lambda^{x+y} & \\ \lambda^{x+y} & & 
    \lambda^{x'+y+z+\alpha} \\ 
    & \lambda^{x+y'+z+\alpha} & \lambda^{x'+y'}
  \end{pmatrix}h,
  \label{NNI2}
\end{equation}
where we have defined $\langle\varphi\rangle\equiv\lambda$ and 
$\langle H_1\rangle\equiv\lambda^\alpha\,$ ($\alpha\geq0$ for a
consistent theory below the fundamental scale).

The eigenvalues of (\ref{NNI2}) are easily 
found; $m_1\sim h\lambda^{x+y-2(z+\alpha)}$, 
$m_2\sim h\lambda^{x+y+2(z+\alpha)}$, 
and $m_3\sim h\lambda^{x'+y'}$. To obtain a hierarchy $m_1\ll m_2$,
the condition $z+\alpha<0$ is naively needed. It should be, however,
noticed that this condition causes a problem that higher-dimensional
operators involving $H_D$ give larger effects and are not necessarily
negligible. We will in the below enumerate possible resolutions to the
problem; (i) to set $z+\alpha>0$, (ii) to take $z<0$ and make
higher-dimensional operators negligible by supersymmetry, (iii) to
tolerate some of higher-dimensional operators, (iv) to impose extra
symmetry, (v) to change the $S_3$ representations of 
fermions, and (vi) to introduce $S_3$-singlet scalars.

\smallskip

(i) If one chooses $z+\alpha>0$, problematic non-renormalizable
operators become irrelevant. This may be the simplest solution in the
viewpoint of $U(1)$ charge assignment. It is found from the matrix
form (\ref{NNI2}) that the mixing angle between the first and
second generations becomes $\pi/2$. Note here that the $S_3$ freedom
cannot be used to reorder the mass eigenvalues because the $S_3$ group
basis has already been fixed such that $H_D$ takes a particular form
of VEV\@. While we know the generation mixing in the quark sector is
small, the $S_3$ realization of the Fritzsch ansatz can be 
consistent (for $z+\alpha>0$), provided that the up and down quark
sectors have almost the same flavor structure. Namely, If both the up
and down sectors employ the matrix form (\ref{NNI2}), the label
exchanging effects are cancelled out between the two sectors, and the
quark mixing angles become small of powers of $\lambda$. Such a
situation is similar to the case of the democratic quark mass
matrices~\cite{S3demo}.

\smallskip

(ii) If one chooses $z<0$, higher-dimensional operators can be
forbidden by holomorphicity in supersymmetric theory. For this being
achieved, it is a simple assignment that all the $U(1)$ charges except
for $H_D$ and $\varphi$ are positive in order not to give any
higher-dimensional terms. Since the doublet scalar $H_D$ now develops
a VEV in the up component, the down component of doublet in the
product $(H_D)^2$ becomes nonzero, and the singlet component 
in $(H_D)^3$ does [see the tensor product (\ref{22c})]. Consequently,
higher-order terms involving the products of $H_D$ might spoil the
first-order analysis, because $H_D$ has a 
negative $U(1)$ charge ($z<0$). In fact, the 1-3, 3-1, and 2-2
elements in the matrix (\ref{NNI}) receive non-negligible
contributions from $(H_D)^2$. The contributions to the 1-3 and 3-1
elements are forbidden by the doublet parity. We however find that
higher-order contribution to the 2-2 element is difficult to be
suppressed as long as the 2-3 and 3-2 matrix elements are allowed. A
simple way to remedy this last problem is to add some extra symmetry,
otherwise to apply the option (iii).

\smallskip

(iii) One, in some sense, negative choice is to abandon the exact
Fritzsch ansatz and put up with some contribution from
non-renormalizable operators. As mentioned above, a non-vanishing 2-2
element often appears even if the parity invariance is imposed. One
can therefore choose as the third option that higher-order terms are
generally forbidden by parity and/or supersymmetry while the 2-2
element is not. For example, in supersymmetric 
approach, the $U(1)$ charges are needed to satisfy the mild
constraints $x+y+z<0$, $x+y'+z\geq0$, $x'+y+z\geq0$, and $z<0$. If
this is the case, we obtain a mass matrix of the 
form (\ref{NNI}) corrected by a nonzero 2-2 element from a
higher-dimensional operator involving $(H_D)^2$. Such a type of mass
texture has been recently discussed~\cite{nonzero22} to be suitable
for solving fermion mass problems including the neutrino physics.

\smallskip

(iv) If one chooses to introduce more additional symmetries, harmful
higher-dimensional operators might be removed. However such an 
operation generally reduces to complicate the models and involve
uncontrollable factors, which make the models unfavorable.

\smallskip

(v) Contrary to the above options (i)--(iv), one can choose to extend
the model to include more fields in other representations of flavor
symmetry. Let us consider an additional Higgs $H_A$ of 
pseudo $S_3$ singlet. The representations of matter fields are
accordingly changed to $(\psi_L{}_2,\psi_L{}_3)+\psi_L{}_1$  
and $(\psi_R{}_2,\psi_R{}_3)+\psi_R{}_1$, i.e.\ the second and third
generations are $S_3$ doublets. It is easily found by 
constructing $S_3$-invariant terms that one still has a mass matrix of
the Fritzsch ansatz, provided that the non-vanishing VEVs are given 
by $\langle H_D\rangle=(0,\langle H_2\rangle)$ 
and $\langle H_A\rangle$. An important difference between this and the
above models is whether the parity symmetry is needed or not to
suppress undesired matrix elements. In the model here, the Fritzsch
ansatz is obtained without imposing any parities. To make the 1-1
element negligible, it is sufficient to take appropriate $U(1)$ charge
assignment, since the first-generation fermions now belong to
different $S_3$ representations from the others, and have 
different $U(1)$ charges. Note that the example here may not be
applied to left-right symmetric cases due to the anti-symmetric matrix
elements generated by a pseudo 
singlet (i.e.\ $M_{23}=-M_{32}$). However asymmetrical forms of mass
texture (zeros) could provide phenomenologically interesting possibility.

\smallskip

(vi) Another choice of additional scalar fields is a singlet 
Higgs $H_S$. A reason to introduce such singlet scalars is to suppress
bare Yukawa couplings [e.g.\ $a$ and $d$ in the 
matrix (\ref{NNI})]. They are induced from other operators
involving $S_3$ singlet scalars with 
non-vanishing $U(1)$ charges. Once the singlet scalars obtain VEVs,
the resultant mass matrix takes the same form as (\ref{NNI}), but
flavor-invariant non-renormalizable operators are different due to the
non-vanishing charges of singlet scalars.\footnote{In case 
that $SU(2)_W$-doublet Higgses belong to non-trivial representations
of $S_3$, such effects of $H_S$ (the suppression of bare Yukawa 
terms) can always be taken into account.} Let $s$ be the $U(1)$ charge
of $H_S$, which must be positive as will be seen below. There are
three possible ways to include the $H_S$ scalar: \ (a) The 3-3 element
comes from an operator involving $H_S$. That needs the charge
conditions $x'+y'<0$ and $x'+y'+s\geq0$. Moreover, for the mass
hierarchy being realized, an additional 
condition $2(z+\alpha)<s+\alpha_s$ is required, 
where $\langle H_S\rangle\equiv\lambda^{\alpha_s}$. It is interesting
that the operators which contribute to the 2-2 element is
automatically suppressed, and one does not need to rely on any
additional symmetries. Furthermore if one 
chooses $s+\alpha_s<\frac{5}{2}(z+\alpha)$ [or $s+\alpha_s<3(z+\alpha)$],  
corrections to the 1-3 and 3-1 [or 1-1] elements become negligibly
small. It can be checked that all the above charge conditions are
easily satisfied. \ (b) In this case, the charge 
conditions $x+y<0$ and $x+y+s\geq0$ are needed for $H_S$ to appear in
the 1-2 and 2-1 elements. A proper mass hierarchy is realized 
when $2(z+\alpha)<s+\alpha_s$, and the 1-3 and 3-1 elements are
negligibly small if $s+\alpha_s<3(z+\alpha)$. Unlike in the 
case (a), these charge conditions do not suppress higher-dimensional
contributions to the 1-1 and 2-2 elements, so some symmetry should be
imposed. \ (c) All the 1-2, 2-1, and 3-3 elements contain 
the $H_S$ field. That 
requires $x+y<0$, $x'+y'<0$, $x+y+s\geq0$, and $x'+y'+s\geq0$. Moreover 
one should take $z+\alpha<s+\alpha_s$ in order for fermion mass
hierarchy to be preserved. With these charge conditions, all the
operators concerned with the 2-2 element are automatically
suppressed. It is, however, found that the 1-1 element generally
receives sizable contribution unless extra symmetry is imposed.

\smallskip

In this way, the nearest-neighbor form of mass matrices, including the
well-known Fritzsch ansatz, is realized in the framework 
of $S_3$ flavor symmetry. Moreover the mass hierarchy among the three
generations is also achieved by incorporating an 
additional $U(1)$ symmetry. An important and non-trivial issue is
whether $U(1)$ charges can be assigned so that they are compatible
with the non-abelian flavor symmetry. We have shown typical examples
of assignments which not only produce mass hierarchy but also suppress
non-renormalizable operators which tend to disturb the mass hierarchy
in the first-order estimation.

%%%%%%%%%%%%%%%%%%%%%%%%%%%%%%%%%%%%%%%%%%%%%%%%%%%%%%%%%%%%%%%%%%%%%%
\subsection{Next-nearest neighbor form}

In most models with $S_3$ flavor symmetry and also in the previous
subsection, the first two light generations are assumed to 
compose $S_3$ doublets. While such an assignment may be favorable to
some phenomenological issues e.g.\ for suppressing flavor-changing
processes, it is not necessarily the unique choice 
for $S_3$ charges. One may easily imagine different, but somewhat
unfamiliar, $S_3$ flavor structures. In the following, we show 
that $S_3$ models with such twisted generations are also relevant to
constructing realistic flavor theory of quarks and leptons. As a
simple example, we here discuss the ansatz for mass texture proposed
in~\cite{Giudice}, which type of texture suggests that the first and
third generation fermions make up $S_3$ doublets.

The procedure is completely parallel to that in the previous
subsection. The key ingredients are the representations of matter and
Higgs fields, the profile of Higgs VEVs, and additional
symmetries. Consider the matter 
representation $(\psi_L{}_1,\psi_L{}_3)+\psi_L{}_2$ 
and $(\psi_R{}_1,\psi_R{}_3)+\psi_R{}_2$, i.e.\ the first and third
generations are $S_3$ doublets. We also introduce an $S_3$-doublet
Higgs scalar $H_D=(H_1,H_2)$, whose down component $H_2$ is assumed to
develop a non-vanishing VEV\@. The electroweak gauge invariance is
implemented by an $SU(2)_W$-doublet scalar $h$, which is a singlet 
of $S_3$. The setup leads to the following form of mass matrix;
\begin{equation}
  \begin{pmatrix}
    & & \!a \\ & \,b & \\ \,a & & \!c\langle H_2\rangle
  \end{pmatrix}h,
  \label{NNNI}
\end{equation}
where 1-2 and 2-1 elements have been suppressed by imposing 
a $Z_2$ parity concerning the second 
generation; $\psi_L{}_2,\psi_R{}_2\to-\psi_L{}_2,-\psi_R{}_2$. Similar
suppression can also be obtained by use of holomorphicity in
supersymmetric theory. At this stage, the 
coefficients $a$, $b$, and $c$ are supposed to contain coupling
constants and scalar VEVs, whose natural sizes 
are ${\cal O}(1)$\@. The texture (\ref{NNNI}) has the form of the
Giudice ansatz~\cite{Giudice} for the up quark mass matrix. The
coefficients $a$ and $b$ should be smaller 
than $c\langle H_2\rangle$ to properly describe mass hierarchy of
fermions. That is realized in the present work by introducing 
flavor $U(1)$ symmetry and a charge-compensating 
scalar $\varphi$. Defining $U(1)$ quantum charges as in the following
table,
\begin{equation}
  \begin{array}{c|ccccccc}
    & \psi_L{}_{1,3} & \psi_L{}_2 & \psi_R{}_{1,3} & \psi_R{}_2 & 
    h & H_D & \varphi \\ \hline
    U(1) & x & x' & y & y' & 0 & z & -1 \\[1mm]
    Z_2 & + & - & + & - & + & + & +
  \end{array}\nonumber
\end{equation}
we have the orders of magnitude of the matrix elements
\begin{equation}
  \begin{pmatrix}
    & & \lambda^{x+y} \\ & \lambda^{x'+y'} & \\ 
    \lambda^{x+y} & & \lambda^{x+y+z+\alpha}\!
  \end{pmatrix}h,
\end{equation}
where $\langle\varphi\rangle\equiv\lambda$, and $\alpha$ parametrizes
the VEV of $S_3$ doublet $\langle H_2\rangle\equiv\lambda^\alpha$. It
is obvious that a charge condition $z+\alpha<0$ is necessary for mass
hierarchy without inducing maximal generation mixing. Notice that this
is similar to the case of the Fritzsch ansatz where higher-dimensional
operators involving the $S_3$ doublet scalar lead to significant
modification of matrix form. Therefore also in the present case, one
can apply the resolutions discussed in the previous subsection to have
natural hierarchy of mass eigenvalues and mixing angles.

%%%%%%%%%%%%%%%%%%%%%%%%%%%%%%%%%%%%%%%%%%%%%%%%%%%%%%%%%%%%%%%%%%%%%%
\subsection{Asymmetric textures}

A more unfamiliar but interesting case is that flavor charges are
asymmetrically assigned to left- and right-handed fermions, which
generally lead to asymmetrical forms of mass textures. In grand
unification schemes, the mass matrix of up-type quarks is often
assumed to be symmetric (exactly speaking, hermitian), which comes
from the fact that, even in the minimal $SU(5)$ model, one-generation
up-type quarks with both chiralities belong to a single multiplet of
unified gauge symmetry. However this is not generally the case for
other Dirac mass matrices. In particular, the present experimental
data suggest that the leptonic flavor mixing is quite un-parallel to
the quark mixing. This asymmetrical observation can be compatible with
quark-lepton unification, if fermion mass textures take asymmetrical 
forms in the generation space~\cite{lopsided}. There are also some
classes of non-hermitian ansatze for quark mass matrices~\cite{non},
which are consistent with the experimental data and cannot be
transformed to the symmetric solutions previously found 
in Ref.~\cite{RRR}. It is therefore worthwhile to investigate the
dynamical realization of asymmetric mass textures with discrete flavor
symmetry. A systematic study of asymmetric mass matrices for the up
and down quarks has recently been performed, particularly paying
attention to the connection to leptonic flavor mixing~\cite{UWY}.

Among various types of viable asymmetric textures, we here present 
an $S_3$ flavor model which predicts the mass matrix proposed
in~\cite{AFM}. This texture is relevant to the neutrino Dirac mass
matrix and is interesting in that large lepton mixing is realized
without any tuning of couplings, if there is a suitable hierarchy
among right-handed neutrino masses. The texture indicates neither that
the first-two generations are in doublet representation nor 
that left- and right-handed fermions have parallel assignments 
of $S_3$ charges. That leads us to consider highly asymmetric flavor 
structure: $(\psi_L{}_1,\psi_L{}_3)+\psi_L{}_2$ 
and $(\psi_R{}_2,\psi_R{}_3)+\psi_R{}_1$. We also 
have $SU(2)_W$-singlet scalars, $H_S$ and $H_D=(H_1,H_2)$, which
are $S_3$ singlet and doublet, respectively. The down 
component $H_2$ of the doublet $H_D$ is assumed to develop a
non-vanishing VEV\@. The electroweak gauge invariance is maintained by
introducing an $SU(2)_W$-doublet standard model Higgs $h$, which is a
singlet of $S_3$. Consequently we obtain the following form of mass
texture:
\begin{equation}
\begin{pmatrix}
  \,a\langle H_2\rangle & d & \\
  & b\langle H_2\rangle & e\langle H_S\rangle \\
  \,e\langle H_S\rangle & & c\langle H_2\rangle
\end{pmatrix}h.
\end{equation}
It is interesting to note that no additional symmetry is required to
eliminate unwanted nonzero matrix elements, contrary to the previous
two examples. The coefficients $a,\cdots,e$ contain coupling constants
whose natural sizes are ${\cal O}(1)$. Splitting the sizes of the
coefficients is easily obtained, for example, by introducing 
flavor $U(1)$ symmetry and a charge-compensating scalar $\varphi$. A
typical $U(1)$ charge assignment is given by
\begin{equation}
  \begin{array}{c|cccccccc}
    & \psi_L{}_{1,3} & \psi_L{}_2 & \psi_R{}_{2,3} & \psi_R{}_1 & 
    h & H_D & H_S & \varphi \\ \hline 
    U(1) & -x & 0 & y & z & 0 & x & 2x & -1
  \end{array}\nonumber
\end{equation}
with $x>y\geq0$ and $z\geq0$, and we obtain the mass texture of the form
\begin{equation}
  \begin{pmatrix}
    \lambda^z & \lambda^z & \\
    & \lambda^{x+y} & \lambda^{x+y} \\
    (\lambda^{x+y}) & & \lambda^y
  \end{pmatrix}h.
  \label{AFMtexture}
\end{equation}
The parameter $\lambda$ denotes the ratio of the 
VEV $\langle\varphi\rangle$ to the fundamental scale. Compared to the
matrix in~\cite{AFM}, a non-vanishing 3-1 element is
generated. However one can easily find that such a tiny entry does not
contribute both to the mass eigenvalues and mixing angles, though it
is obliged to be included to respect the flavor symmetry.

It was discussed~\cite{AFM} that the 
texture (\ref{AFMtexture}) (without the 3-1 element) could be derived
by making use of more than two continuous flavor symmetries. By
contrast, the present approach with discrete flavor symmetry seems
somewhat simpler. The texture form is totally controlled by a 
single $S_3$ symmetry, and then, the mass hierarchy is determined by
extra assumption (here the $U(1)$ flavor symmetry).

Finally let us comment on the corrections from non-renormalizable
operators. It is found that the above charge assignment suppresses
the contribution of higher-dimensional operators to 
the 1-3 and 3-2 matrix elements, and the texture zeros are not
destabilized. On the other hand, the correction to the 2-1 element
could be as large as the other components in the second 
rows, i.e.\ ${\cal O}(\lambda^{x+y})$. This result itself is not a
disaster but the above zero texture should be modified. That is
similar to the cases in the previous subsections where
higher-dimensional operators involving $H_D$ could give significant
effects on the first-order approximation. Therefore one is able to
apply similar resolutions to remedy the problem. For example, if 
the $S_3$ doublet scalar $H_D$ has a non-trivial charge of some other
symmetry, any problematic deformation of mass matrix is completely
avoidable.

%%%%%%%%%%%%%%%%%%%%%%%%%%%%%%%%%%%%%%%%%%%%%%%%%%%%%%%%%%%%%%%%%%%%%%
\subsection{Missing a right-handed neutrino}
\label{missing}

Referring to the $S_3$ algebra presented in Section~\ref{S3group}, the
maximal irreducible representation is doublet. The triplet
representation is reducible, while it may be suitable to describe
three-generation fermions. Two fermions with the same standard model
charges are also well handled with $S_3$ symmetry. While we have
already known the existence of three repetitions of the standard model
fermions, there is an idea that only two right-handed neutrinos are
effectively included~\cite{2nuR}. In this framework, a missing
right-handed neutrino might be thought to decouple at a superheavy
scale and regarded as a flavor singlet.\footnote{Such a singlet should
be $1_{\rm S}$ to preserve $S_3$ symmetry in low-energy regime. The 
existence of the third right-handed neutrino would be needed to
reconcile the lepton sector to the quark one in high-energy
unification schemes.} The remaining two right-handed neutrinos are
hence treated as a doublet. In this subsection, we present a model
with three left-handed and two right-handed neutrinos in the framework
of $S_3$ flavor symmetry.

We would like here to stress that another important issue of flavor
symmetry is to simultaneously control different sectors of flavor. For
example, physical quark mixing angles are determined with a conspiracy
of the up and down mass matrices, and low-energy Majorana neutrino
masses are derived from Dirac and right-handed Majorana masses with
the seesaw mechanism~\cite{seesaw}. In the following, we illustrate
simultaneous treatment of two different types of mass 
textures, i.e.\ the Dirac and Majorana masses of neutrinos. A more
realistic flavor model, including quarks and charged leptons, will be
discussed in the next section.

Among various models with two right-handed neutrinos, we here focus
on the texture ansatz proposed in~\cite{FGY} and its realization
with $S_3$ flavor symmetry in the neutrino sector. The model contains
three left-handed leptons $\psi_L{}_{1,2,3}$ and two right-handed 
ones $\psi_R{}_{1,2}$. We assign the flavor charges to these fermions
such that ($\psi_L{}_1,\psi_L{}_2$) and ($\psi_R{}_1,\psi_R{}_2$) are
$S_3$ doublets and $\psi_L{}_3$ a singlet (either of the $S_3$ singlet
representations is possible). As for the Higgs sector, at least two
types of scalars are introduced for controlling two different sectors
of mass textures, though not necessarily required. We take, as an
example, two scalar fields which are $S_3$ doublets and the standard
gauge singlets; $H_D=(H_1,H_2)$ and $H_D'=(H_1',H_2')$, whose VEVs
generate Dirac and Majorana masses, respectively. For compensating
electroweak gauge invariance, a usual $SU(2)_W$-doublet Higgs $h$ is
also included. A more economical choice might be to 
have $SU(2)_W$-doublet and $S_3$-doublet scalar fields. In this 
case, however, there may exist some problems, as discussed 
before, that gauge coupling unification must be non-trivially realized
and large rates of flavor-changing processes generally spoil the models.

The $S_3$ invariant terms induce the neutrino Dirac mass $M_\nu$ and
the right-handed Majorana mass $M_R$, which read from the generic
expressions (\ref{MD}) and (\ref{MR}),
\begin{align}
  M_\nu &= \begin{pmatrix}
  & a\langle H_2\rangle+b\langle H_1^*\rangle & 
  c\langle H_1\rangle+d\langle H_2^*\rangle \\
  a\langle H_1\rangle+b\langle H_2^*\rangle & & 
  c\langle H_2\rangle+d\langle H_1^*\rangle
  \end{pmatrix}h, \\[1mm]
  M_R &= \begin{pmatrix}
  e\langle H_1'\rangle+f\langle H_2'{}^*\rangle & \\
  & e\langle H_2'\rangle+f\langle H_1'{}^*\rangle
  \end{pmatrix},
\end{align}
where $a,\cdots,f$ are the coupling constants. Thus we have obtained
the textures discussed in~\cite{FGY}, if the generation indices are
properly exchanged while physical consequences are unchanged (except
for a sign reversion of as yet unobserved leptonic CP-violating
quantity). The bi-large generation mixing in the neutrino sector is
naturally established, provided that $(M_\nu)_{12}\simeq (M_\nu)_{13}$ 
and $(M_\nu)_{21}\simeq (M_\nu)_{23}$, and that there is a little
hierarchy between these two combinations or between the mass
eigenvalues of $M_R$~\cite{FGY}.

It is rather straightforward to incorporate supersymmetry into the
above picture. Every field is promoted to a superfield which is in the
same representations of the standard model gauge and flavor
symmetries. The Dirac and Majorana mass terms are described by
superpotential. Notice that the right-handed 
neutrinos $\psi_R{}_i$ are embedded into the corresponding superfields
in the form $\psi_R^{\,c}{}_i$. As we explained in Section~\ref{inv},
this embedding modifies the tensor product and slightly changes the
texture form. We thus find that the $S_3$-invariant superpotential
induces the mass matrices
\begin{equation}
  M_\nu \,=\, \begin{pmatrix}
  a\langle H_1\rangle & & b\langle H_2\rangle \\
  & a\langle H_2\rangle & b\langle H_1\rangle
  \end{pmatrix}, \qquad
  M_R \,=\, \begin{pmatrix}
  c\langle H_1'\rangle & \\ & c\langle H_2'\rangle
  \end{pmatrix}.
\end{equation}
Again the neutrino mass textures presented in~\cite{FGY} are
dynamically realized in a supersymmetric $S_3$ model. It is easy to
see that the coupling constants and scalar VEVs can be appropriately
chosen for the neutrino physics without any fine tuning.

%%%%%%%%%%%%%%%%%%%%%%%%%%%%%%%%%%%%%%%%%%%%%%%%%%%%%%%%%%%%%%%%%%%%%%
\section{Grand unification with flavor $\boldsymbol{S_3}$}

Based on the above prescription for mass texture, in this section, we
present a grand unified model of quarks and leptons with $S_3$ flavor
symmetry. It is the most important point of flavor symmetry that any
matrix form from flavor symmetry is valid only in the case that there
are more than two sectors governed by the symmetry. A well-known
example is the up and down quark mass textures whose forms are
simultaneously altered by flavor rotation of $SU(2)_W$-doublet
quarks. In other words, any realization of zero textures for a single
sector cannot be physically distinguished from other freely-rotated,
generally complicated, matrix forms. In the previous subsection, we
discussed a model in which two types of neutrino mass matrices are
controlled in the same fashion. To include the charged lepton sector
is straightforward, referring to the general discussion 
of $S_3$-invariant matrix forms. Towards realistic flavor models of
quarks and leptons, we here construct an illustrative grand unified
model with $S_3$ flavor symmetry. The model has asymmetrical forms of
mass textures due to non-trivial assignment of flavor charges, which
particularly leads to bi-large lepton mixing. While the model
presented here is not so complete, it suggests the validity of
discrete symmetry for flavor physics and would give a large step to
understand the origin of flavor.

Let us consider $SU(5)$ as a minimal candidate for unified gauge
group. For matter Yukawa couplings, we introduce as usual the
three-generation 
fermions, $\psi_i(10)$, $\chi_i(5^*)$, $\nu_i(1)$ ($i=1,2,3$) and two
scalars, $h_u(5)$ and $h_d(5^*)$. Needless to say, one needs some
different types of scalar fields to break the unified gauge symmetry,
but they are generally irrelevant to the Yukawa sector. We define the
flavor $S_3$ charges of matter and scalar fields so that the
combinations ($\psi_1,\psi_2$), ($\chi_1,\chi_3$) and ($\nu_1,\nu_2$)
are $S_3$ doublets and the others in 
the $1_{\rm S}$ representation. To implement the $S_3$ invariance, a
doublet and a singlet scalars are also 
introduced, called $H_D$ and $H_S$, respectively. The total field
content is listed in the table below. We assume that $S_3$ is broken
by the scalar VEVs; $\langle H_D\rangle=(\langle H_1\rangle,0)$ 
and $\langle H_S\rangle$, which are of ${\cal O}(1)$ and can be the
result of analyzing symmetry-invariant scalar potential without fine
tuning.
\begin{equation}
\begin{array}{c|ccccccccccc}
& \psi_{1,2} & \psi_3 & \chi_{1,3} & \chi_2 & \nu_{1,2} & \nu_3 & 
h_u & h_d & H_D & H_S & \varphi \\ \hline
SU(5) & 10 & 10 & 5^* & 5^* & 1 & 1 & 5 & 5^* & 1 & 1 & 1 \\[0.5mm]
S_3 & 2 & 1 & 2 & 1 & 2 & 1 & 1 & 1 & 2 & 1 & 1 \\[0.5mm]
U(1) & 4 & 0 & 2 & 0 & x+2 & y & -2 & -2 & 0 & 2 & -1
\end{array}\nonumber
\end{equation}

While the right-handed neutrinos are assigned to 
the $S_3$ representations $(\nu_1,\nu_2)+\nu_3$, that is not essential
as long as the low-energy left-handed Majorana masses are
concerned. It is because, in the seesaw mechanism, arbitrary rotations
of right-handed neutrinos can be inserted while preserving left-handed
Majorana mass matrix. In fact, the $S_3$ flavor charges (i.e.\ the
flavor basis) of $\nu_i$ turn out to be fixed, once some other
dynamics than $S_3$ is incorporated.

In addition to these, we have a $U(1)$ (or $Z_N$) symmetry to produce
natural hierarchies among matrix elements (and also to help to 
discriminate $\nu_i$). A key of the mechanism is the existence of a
scalar field $\varphi$ with a non-vanishing VEV, which is a bit
smaller than the fundamental scale of the theory. The 
whole $S_3$ representations and $U(1)$ quantum numbers are shown in
the table ($x\geq0$ and $2>y\geq0$ for phenomenological requirements).

Let us first discuss the Majorana masses of right-handed
neutrinos. They are given by the following form of $S_3$-invariant 
mass operator $\overline{\nu^c_i}(M_R)_{ij}\nu_j$:
\begin{equation}
  M_R \,=\, \begin{pmatrix}
  & M & \\
  M & & \\
  & & M'
  \end{pmatrix},
  \label{MRGUT}
\end{equation}
where the mass parameters $M$ and $M'$ are generally free. They
contain some compound factors such as suppressions by 
flavor $U(1)$ invariance, but such suppression factors are cancelled
out via the seesaw mechanism and irrelevant for low-energy
quantity. The orders of magnitudes of $M$ and $M'$ would be determined
by experimental data. Theoretically this seems natural 
since $M_R$ cannot be directly induced by $h_{u,d}$, nor 
come from $H_{D,S}$, as the Majorana mass term violates lepton number
symmetry. Otherwise one easily obtains $M_R$ by assuming VEVs 
of $S_3$-singlet scalar fields with lepton numbers. In any case, the
most important point here is that a single $S_3$ flavor symmetry
controls both Dirac and Majorana mass textures simultaneously.

We also find that the $S_3$ and $U(1)$ flavor invariance leads to the
following Dirac mass textures $M_{u,d,e,\nu}$ for quarks and leptons:
\begin{gather}
M_u \,=\, 
\begin{pmatrix}
  \lambda^6 & \lambda^6 & \\
  \lambda^6 & & \lambda^2 \\
  & \lambda^2 & 1
  \end{pmatrix}h_u, \qquad
M_d \,=\,
\begin{pmatrix}
  \lambda^4 & & \lambda^4 \\
  \lambda^4 & \lambda^2 & \\
  & & 1
\end{pmatrix}h_d, \\[2mm]
M_e \,=\, M_d^{\rm T}, \qquad\qquad
M_\nu \,=\,
\begin{pmatrix}
\lambda^{x+2} & & \lambda^{x+2} \\
\lambda^{x+2} & \lambda^x & \\
& \lambda^y & \lambda^y
\end{pmatrix}h_u,
\end{gather}
where we have neglected coupling constants and defined the expansion
parameter $\lambda$ as the ratio of $\langle\varphi\rangle$ to the
fundamental scale. Our first prediction obtained from the above 
textures is that the VEV ratio of two $SU(2)_W$ doublets should be
large, that is, $h_u/h_d\sim m_t/m_b\sim 60$. Secondly, we find the
Majorana mass matrix $M_L$ of left-handed neutrinos derived from the
seesaw formula
\begin{equation}
  M_L \,=\, \begin{pmatrix}
    & \epsilon & \\
    \epsilon & 1 & 1+\epsilon \\
    & 1+\epsilon & 1
  \end{pmatrix}\frac{\lambda^{2y}h_u^2}{M'}.
  \label{ML}
\end{equation}
Here the parameter $\epsilon$ has been defined 
as $\epsilon\equiv\lambda^{2x-2y+2}(M'/M)$. The induced matrix
elements in the 1-1, 1-3 and 3-1 positions are negligibly small and
have been dropped in the expression (\ref{ML}). It is interesting to
note that all the symbols ``1'' in the matrix (\ref{ML}) are 
exactly 1 due to the seesaw mechanism, and consequently, the
lower-right $2\times 2$ sub-matrix has a reduced determinant 
of ${\cal O}(\epsilon)$, not ${\cal O}(1)$. For a  
small $\epsilon$ ($\sim\lambda^{1-2}$), the matrix (\ref{ML})
automatically realizes the bi-large generation mixing with
hierarchical neutrino spectrum. The model also predicts a small lepton
mixing of ${\cal O}(\epsilon)$ between the first and third
generations. The planned improvements in the sensitivity to such a
small angle are expected to 
reach ${\cal O}(10^{-2})$~\cite{futurenuexp}, and the above model will
be testable in near future. On the other hand, the neutrinoless double
beta decay~\cite{nulessbb} cannot be observed due to the negligible
value of $(M_L)_{11}$.

It is found in the above Dirac mass matrices that some distorted
values are induced for light generations, while the coefficients of
matrix elements could be appropriately chosen for reproducing the
observed values of fermion masses and mixing angles. We checked that,
for the present simple field content, the flavor charge assignment 
of $S_3$ and $U(1)$ symmetries is uniquely determined for realizing
the two conditions that (i) the right-handed neutrino 
masses $M_R$ come from the mass operator (\ref{MRGUT}) and (ii) the
rank of $2\times 2$ sub-matrix of $M_L$ is reduced via the seesaw
mechanism without tuning of couplings. Therefore an improvement of the
above toy model is made by extending it to include additional scalar
fields and modifying the texture forms. For example, the equivalence
of the down and charged-lepton mass spectrum may be split by making
use of Higgs fields in higher-dimensional representations. Another
simple extension is to change the $S_3$ group basis, e.g.\ to the real
basis. This might be a possible amelioration because the texture
analysis and its physical consequences depend on the group basis of
flavor symmetry, as explained before. We leave these issues to future
investigation.

%%%%%%%%%%%%%%%%%%%%%%%%%%%%%%%%%%%%%%%%%%%%%%%%%%%%%%%%%%%%%%%%%%%%%%
\section{Analysis of scalar potentials}

We have so far discussed mass textures of fermions in the framework 
of $S_3$ flavor symmetry, where the $S_3$ invariance is violated only
by scalar VEVs. In particular, the assumption that either of
components in a doublet has a vanishing VEV leads to texture forms for
quark and lepton mass matrices, which result in various types of
phenomenologically viable flavor models. In this section, we present,
as an existence proof, the analysis of scalar potentials which
generate desired forms of symmetry-breaking VEVs. We also discuss
model parameters and symmetry for obtaining such VEVs. It should be
noted that, unlike in the case of continuous flavor symmetry, a VEV of
doublet cannot be transformed to an arbitrary form by $S_3$ rotations
because of the discreteness of group 
operations (see Table~\ref{matrix}). Even if there is a case that such
rotation is useful, it is physically meaningful only when other
symmetries in the theory do not commute with $S_3$. This is not
necessarily satisfied, for example, in the presence of $U(1)$ flavor
symmetry.

%%%%%%%%%%%%%%%%%%%%%%%%%%%%%%%%%%%%%%%%%%%%%%%%%%%%%%%%%%%%%%%%%%%%%%
\subsection{The general case}

We consider a single $S_3$ doublet scalar $H=(H_1,H_2)$ and analyze
the structure of its generic scalar potential. The potential analysis
becomes more involved if there are some numbers of scalar
fields. However since we have shown that mass textures of quarks and
leptons can be derived from a single doublet scalar, the analysis for
one doublet would provide a step towards more practical cases with
more than one doublets. Furthermore, when several scalar fields have
VEVs of enough separated scales, they can be independently analyzed if
the most generic terms are taken into account in scalar potentials.

The most general and renormalizable scalar potential for a 
doublet $H=(H_1,H_2)$ is given by
\begin{eqnarray}
  V(H_1,H_2) &=& \mu^2 (H_1^\dagger H_1+H_2^\dagger H_2)
  +\lambda\big[(H_1^\dagger H_1)^2+(H_2^\dagger H_2)^2\big] 
  \nonumber \\[1mm] 
  && +\big[\rho^2H_1H_2 +\alpha(H_1^3+H_2^3)+
  \beta(H_1^2H_2^\dagger+H_2^2H_1^\dagger) 
  +{\rm h.c.}\big] \nonumber \\[1mm]
  && +\kappa H_1^\dagger H_1H_2^\dagger H_2
  +\big[\zeta H_1^2H_2^2+\eta(H_1^\dagger H_1^2H_2
  +H_2^\dagger H_2^2H_1) +{\rm h.c.}\big]
  \label{nonSUSYV}
\end{eqnarray}
with respect to the $S_3$ invariance in the complex basis. The
minimization of the potential leads to the following two conditions:
\begin{eqnarray}
  0\,=\,\frac{\partial V}{\partial H_1^\dagger} &=& 
  \mu^2H_1 +2\lambda H_1^\dagger H_1^2 +\rho^{2*}H_2^\dagger
  +3\alpha^*H_1^{\dagger 2} +\beta H_2^2 
  +2\beta^*H_1^\dagger H_2 +\kappa H_1H_2^\dagger H_2 \nonumber \\
  && \qquad +2\zeta^*H_1^\dagger H_2^{\dagger 2} 
  +\eta H_1^2H_2 +\eta^*(2H_1^\dagger H_1H_2^\dagger
  +H_2^{\dagger 2}H_2), \label{V1} \\[1mm]
  0\,=\,\frac{\partial V}{\partial H_2^\dagger} &=& 
  \mu^2H_2 +2\lambda H_2^\dagger H_2^2 +\rho^{2*}H_1^\dagger
  +3\alpha^*H_2^{\dagger 2} +\beta H_1^2 
  +2\beta^*H_1H_2^\dagger +\kappa H_1^\dagger H_1H_2 \nonumber \\
  && \qquad +2\zeta^*H_1^{\dagger 2}H_2^\dagger 
  +\eta H_1H_2^2 +\eta^*(H_1^{\dagger 2}H_1
  +2H_1^\dagger H_2^\dagger H_2). \label{V2}
\end{eqnarray}
The minimum is ensured by examining whether the $2\times 2$ matrix 
$\partial^2 V/\partial H_i\partial H_j^\dagger$ ($i,j=1,2$) has
positive eigenvalues at the extreme. Generally, the stationary 
conditions (\ref{V1}) and (\ref{V2}) are satisfied at various
points in the field space. The origin $H_1=H_2=0$ is a solution of
these two condition, but at this point, the $S_3$ flavor symmetry is
unbroken and the result is unrealistic. Parallel to the well-known
electroweak symmetry breaking, one easily avoids this trivial solution
by assuming a negative value of $\mu^2$, which in turn makes the
origin being a locally maximum point and 
destabilized.\footnote{Exactly speaking, if $\mu^2<|\rho^2|$, the 
origin of field space is destabilized, but only along with limited
directions.} A possible minimum we are interested in is that either of
VEVs in a doublet scalar is 
vanishing, namely $H_1\neq0$ and $H_2=0$, or vice 
versa. For $H_2=0$, the stationary 
conditions (\ref{V1}) and (\ref{V2}) become
\begin{eqnarray}
\mu^2H_1+3\alpha^*H_1^{\dagger 2} +2\lambda H_1^\dagger H_1^2 &=& 0, \\
\rho^2H_1+\beta^*H_1^{\dagger 2} +\eta H_1^\dagger H_1^2 &=& 0.
\end{eqnarray}
It is found that the existence of non-trivial (at least local) minimum 
with $H_1\neq0$ requires a set of conditions for the coupling
constants. For example, $\rho=\beta=\eta=0$ or $\mu^2/\rho^2=
3\alpha/\beta=2\lambda/\eta$, and so on. The former condition could 
easily be obtained if the scalar has some other quantum charges
than $S_3$. There exist various possibilities of coupling constants
for stabilizing the scalar potential at non-trivial minimum. It is an
intriguing task to find the conditions of couplings which are
naturally realized with symmetries. Finally we comment on the VEV 
form $H_1\neq0$, $H_2\neq 0$. Such a form of doublet VEVs is needed to
achieve the texture form discussed in Section~\ref{missing}. If the
coupling constants in the above scalar potential have the same order
of magnitude (in the unit of fundamental scale), two 
VEVs $H_1$ and $H_2$ are on a similar order and generally have
different values. In this case, it is difficult to write down the
generic solution to (\ref{V1}) and (\ref{V2}) for the potential
minimum. We numerically investigated that there indeed exists the
minimum of the potential (\ref{nonSUSYV}). As an example, a set of
natural values of coupling 
constants, $\mu=\rho=\alpha=\beta\equiv m$ 
and $\lambda=\kappa=\zeta=\eta=1$, leads to the stationary 
point $H_1=0.61m$ and $H_2=-1.62m$. It can easily be checked that this
point corresponds to a potential minimum by analyzing the second order
derivatives of the potential. It is also interesting to explore the
region of coupling constants for which such a form of doublet VEVs
appears.

%%%%%%%%%%%%%%%%%%%%%%%%%%%%%%%%%%%%%%%%%%%%%%%%%%%%%%%%%%%%%%%%%%%%%%
\subsection{Supersymmetric case}

The analysis is similarly performed in supersymmetric theory. For 
an $S_3$-doublet superfield $H=(H_1,H_2)$, the symmetry-invariant
superpotential $W$ is given by
\begin{equation}
  W \,=\, mH_1H_2 +\frac{y}{3}(H_1^3+H_2^3) +\frac{w}{2}H_1^2H_2^2
  +xH_1H_2(H_1^3+H_2^3) +\frac{z_1}{6}(H_1^6+H_2^6)
  +\frac{z_2}{3}H_1^3H_2^3,
  \label{SUSYV}
\end{equation}
where $m$, $y$, $\cdots$, $z_1$, $z_2$ are the coupling constants and
we have included the most general operators with mass dimensions up to
seven. In this subsection, we assume that supersymmetry is unbroken,
that is, supersymmetry-breaking soft terms do not fix the breaking
scale of flavor symmetry. The scalar potential is then given by
\begin{eqnarray}
  V(H_1,H_2) &=& \Big|mH_1+yH_2^2+wH_1^2H_2+x(4H_1H_2^3+H_1^4)
    +z_1H_2^5+z_2H_1^3H_2^2\Big|^2 \nonumber \\
  && \quad +\Big|mH_2+yH_1^2+wH_1H_2^2+x(4H_1^3H_2+H_2^4)
    +z_1H_1^5+z_2H_1^2H_2^3\Big|^2. \quad
\end{eqnarray}
Comparing to the general case (\ref{nonSUSYV}), we have the 
relations from supersymmetry:
\begin{equation}
  \mu^2=|m|^2,\quad \lambda=|y|^2,\quad \beta=m^*y,\quad \eta=m^*w,
  \quad \rho^2=\alpha=\kappa=\zeta=0.
\end{equation}

We here focus on some simple cases obtained by introducing discrete
symmetry on the superpotential. The vacuum is where one of the VEVs of
doublet is vanishing, e.g.\ $H=(H_1,0)$. Let us consider 
a $Z_3$ symmetry of which the superfield $H$ has a charge +1. In this
case, it permits only the $y$, $z_1$ and $z_2$ terms in the
superpotential (\ref{SUSYV}). As in the general case without
supersymmetry, the origin $H_1=H_2=0$ is a trivial solution without
flavor symmetry breaking. We here assume for simplicity that the
origin is destabilized if 
supersymmetry-breaking (and $S_3$-breaking) soft scalar terms are
included. We then find a potential minimum where the scalars develop
their VEVs of the form $H=\big((-y/z_1)^{1/3},0\big)$. At this 
vacuum, a vanishing VEV can give rise to mass textures (zeros) for
quarks and leptons as explained before. Another simple example is a
discrete $R$ symmetry $Z_{3R}$ under which the superfield $H$ has a
charge +1. In this case, only the $m$ and $x$ terms are allowed in the
superpotential (\ref{SUSYV}). Similarly assuming that the origin does
not correspond to the minimum, we find a vacuum at which the doublet
has the one-sided form of VEVs $H=\big((-m/x)^{1/3},0\big)$. It may be
interesting to note that the $Z_{3(R)}$ symmetry is suitable for
discussing trilinear Yukawa terms and also for suppressing
non-renormalizable operators involving more than 
one $S_3$ doublets. We finally comment that the analysis becomes more
involved if higher-dimensional operators in K\"ahler terms were
included.

%%%%%%%%%%%%%%%%%%%%%%%%%%%%%%%%%%%%%%%%%%%%%%%%%%%%%%%%%%%%%%%%%%%%%%
\section{Summary and discussions}

Discrete flavor symmetry is a powerful instrument in classifying and
constructing mass matrix forms of quarks and leptons. It is greatly
anticipated that such discrete symmetry is practically inherited from
continuous symmetries in more fundamental theory in high-energy
regime. In this paper, we focus on the $S_3$ group as a promising
candidate of such symmetry. The $S_3$ is the smallest non-abelian
discrete symmetry and also has a simple geometrical
interpretation. After discussing some fundamental issues of 
the $S_3$ group (representations, tensor products and group basis
dependence), we have presented the general forms of $3\times3$ mass
matrices which are derived from various types of $S_3$ theories. Based
on that prescription, we have performed the construction of particular
ansatze of mass textures which have been often discussed in the
literature. That includes asymmetrical matrices in generation space,
which would be appropriate to discussing the leptonic flavor mixing.

Furthermore, discrete flavor symmetry can control the structure of
mass matrices (texture) in dynamical ways. In particular, we have
shown that vanishing matrix elements (texture zeros) are dynamically
realized in the vacuum of scalar potential. It is however noted that
texture zeros themselves do not give any explanation for fermion mass
hierarchies. The observed values of masses and mixing angles are
obtained in our scheme by introducing a $U(1)$ or $Z_N$ symmetry. A
non-trivial issue here is whether abelian charges can be assigned so
that they are compatible with non-abelian flavor symmetry. We have
described typical examples of charge assignments which not only
produce mass hierarchy among the generations but also suppress
higher-dimensional operators which tend to disturb mass hierarchies in
the first-order estimation. Unlike previous approaches with 
the $S_3$ group, we do not assume hierarchical coupling constants nor
sequential breaking of flavor symmetry. In realistic flavor models for
quarks and leptons, it is important that several different sectors are
simultaneously controlled by flavor symmetry. As an illustrative
example, a grand unified model has been constructed with a 
single $S_3$ symmetry. Finally, we have also analyzed the scalar
potentials for $S_3$-doublet scalars, whose form of VEV is a key
ingredient of our approach.

In this paper (especially below Section~\ref{massS3}), we have focused
on the models with a single $S_3$ symmetry. It may be an interesting
task to explore other types of flavor 
extensions, e.g.\ $S_3{}_L\times S_3{}_R$ and other discrete
symmetries larger than $S_3$. They would generally lead to different
types of mass textures and physical consequences such as
flavor-violating processes and their characteristic experimental
signatures. More detailed study, including these issues, is left to
future investigations.

%%%%%%%%%%%%%%%%%%%%%%%%%%%%%%%%%%%%%%%%%%%%%%%%%%%%%%%%%%%%%%%%%%%%%%
\bigskip
\subsection*{Acknowledgments}

We would like to thank M.~Tanimoto for helpful discussion. This work
is supported by scientific grants from the Ministry of 
Education (No.~16028214, 16540258, 17740150) and by grant-in-aid for
the scientific research on priority area "Progress in elementary
particle physics of the 21st century through discoveries of Higgs
boson and supersymmetry" (No.~16081209).

%%%%%%%%%%%%%%%%%%%%%%%%%%%%%%%%%%%%%%%%%%%%%%%%%%%%%%%%%%%%%%%%%%%%%%


\newpage
\begin{thebibliography}{99}
\bibitem{Fritzsch}
H.~Fritzsch,
%``Weak Interaction Mixing In The Six - Quark Theory,''
{\sl Phys.~Lett.} {\bf B73} (1978) 317.
%%CITATION = PHLTA,B73,317;%%

\bibitem{GJ}
H.~Georgi and C.~Jarlskog,
%``A New Lepton - Quark Mass Relation In A Unified Theory,''
{\sl Phys.~Lett.} {\bf B86} (1979) 297.
%%CITATION = PHLTA,B86,297;%%

\bibitem{Giudice}
G.F.~Giudice,
%``A New ansatz for quark and lepton mass matrices,''
{\sl Mod.~Phys.~Lett.} {\bf A7} (1992) 2429.
%[arXiv:hep-ph/9204215].
%%CITATION = HEP-PH 9204215;%%

\bibitem{RRR}
P.~Ramond, R.G.~Roberts and G.G.~Ross,
%``Stitching the Yukawa quilt,''
{\sl Nucl.~Phys.} {\bf B406} (1993) 19.
%[arXiv:hep-ph/9303320].
%%CITATION = HEP-PH 9303320;%%

\bibitem{FGM}
P.H.~Frampton, S.L.~Glashow and D.~Marfatia,
%``Zeroes of the neutrino mass matrix,''
{\sl Phys.~Lett.} {\bf B536} (2002) 79.
%[arXiv:hep-ph/0201008].
%%CITATION = HEP-PH 0201008;%%

\bibitem{atmos}
Y.~Fukuda {\it et al.},
%``Measurement of a small atmospheric nu/mu / nu/e ratio,''
{\sl Phys.~Lett.} {\bf B433} (1998) 9;
%[arXiv:hep-ex/9803006].
%%CITATION = HEP-EX 9803006;%%
%Y.~Fukuda {\it et al.}  [Super-Kamiokande Collaboration],
%``Study of the atmospheric neutrino flux in the multi-GeV energy
%range,''
{\it ibid.} {\bf B436} (1998) 33;
%[arXiv:hep-ex/9805006].
%%CITATION = HEP-EX 9805006;%%
%Y.~Fukuda {\it et al.}  [Super-Kamiokande Collaboration],
%``Evidence for oscillation of atmospheric neutrinos,''
{\sl Phys.~Rev.~Lett.} {\bf 81} (1998) 1562;
%[arXiv:hep-ex/9807003].
%%CITATION = HEP-EX 9807003;%%
%Y.~Fukuda {\it et al.}  [Super-Kamiokande Collaboration],
%``Measurement of the flux and zenith-angle distribution of upward
%through-going muons by Super-Kamiokande,''
{\it ibid.} {\bf 82} (1999) 2644;
%[arXiv:hep-ex/9812014].
%%CITATION = HEP-EX 9812014;%%
%S.~Fukuda {\it et al.}  [Super-Kamiokande Collaboration],
%``Tau neutrinos favored over sterile neutrinos in atmospheric muon  
%neutrino oscillations,''
{\it ibid.} {\bf 85} (2000) 3999.
%[arXiv:hep-ex/0009001].
%%CITATION = HEP-EX 0009001;%%

\bibitem{solar}
S.~Fukuda {\it et al.},
%``Solar B-8 and he p neutrino measurements from 1258 days of
%Super-Kamiokande data,''
{\sl Phys.~Rev.~Lett.} {\bf 86} (2001) 5651;
%[arXiv:hep-ex/0103032].
%%CITATION = HEP-EX 0103032;%%
%S.~Fukuda {\it et al.}  [Super-Kamiokande Collaboration],
%``Constraints on neutrino oscillations using 1258 days of
%Super-Kamiokande solar neutrino data,''
{\it ibid.} {\bf 86} (2001) 5656;
%[arXiv:hep-ex/0103033].
%%CITATION = HEP-EX 0103033;%%
%S.~Fukuda {\it et al.}  [Super-Kamiokande Collaboration],
%``Determination of solar neutrino oscillation parameters using 1496
%days of Super-Kamiokande-I data,''
{\sl Phys.~Lett.} {\bf B539} (2002) 179;
%[arXiv:hep-ex/0205075].
%%CITATION = HEP-EX 0205075;%%
Q.R.~Ahmad {\it et al.},
%``Direct evidence for neutrino flavor transformation from
%neutral-current interactions in the Sudbury Neutrino Observatory,''
{\sl Phys.~Rev.~Lett.} {\bf 89} (2002) 011301;
%[arXiv:nucl-ex/0204008].
%%CITATION = NUCL-EX 0204008;%%
%Q.R.~Ahmad {\it et al.}  [SNO Collaboration],
%``Measurement of day and night neutrino energy spectra at SNO and 
%constraints on neutrino mixing parameters,''
{\it ibid.} {\bf 89} (2002) 011302;
%[arXiv:nucl-ex/0204009].
%%CITATION = NUCL-EX 0204009;%%
K.~Eguchi {\it et al.},
%``First results from KamLAND: Evidence for reactor anti-neutrino
%disappearance,''
{\sl Phys.~Rev.~Lett.} {\bf 90} (2003) 021802;
%[arXiv:hep-ex/0212021].
%%CITATION = HEP-EX 0212021;%%
T.~Araki {\it et al.},
%``Measurement of neutrino oscillation with KamLAND: Evidence of
%spectral distortion,''
{\sl Phys.~Rev.~Lett.} {\bf 94} (2005) 081801.
%[arXiv:hep-ex/0406035].
%%CITATION = HEP-EX 0406035;%%

\bibitem{sneudom}
S.F.~King,
%``Atmospheric and solar neutrinos with a heavy singlet,''
{\sl Phys.~Lett.} {\bf B439} (1998) 350;
%[arXiv:hep-ph/9806440].
%%CITATION = HEP-PH 9806440;%%
%S.F.~King,
%``Constructing the large mixing angle MNS matrix in see-saw models
%with right-handed neutrino dominance,''
{\sl JHEP} {\bf 0209} (2002) 011;
%[arXiv:hep-ph/0204360].
%%CITATION = HEP-PH 0204360;%%
S.~Davidson and S.F.~King,
%``Bi-maximal neutrino mixing in the MSSM with a single right-handed
%neutrino,''
{\sl Phys.~Lett.} {\bf B445} (1998) 191.
%[arXiv:hep-ph/9808296].
%%CITATION = HEP-PH 9808296;%%

\bibitem{Rparity}
A.~Santamaria and J.W.F.~Valle,
%``Spontaneous R Parity Violation In Supersymmetry: A Model For Solar
%Neutrino Oscillations,''
{\sl Phys.~Lett.} {\bf B195} (1987) 423;
%%CITATION = PHLTA,B195,423;%%
R.~Hempfling,
%``Neutrino Masses and Mixing Angles in SUSY-GUT Theories with
%explicit R-Parity Breaking,''
{\sl Nucl.~Phys.} {\bf B478} (1996) 3;
%[arXiv:hep-ph/9511288].
%%CITATION = HEP-PH 9511288;%%
F.M.~Borzumati, Y.~Grossman, E.~Nardi and Y.~Nir,
%``Neutrino masses and mixing in supersymmetric models without R
%parity,''
{\sl Phys.~Lett.} {\bf B384} (1996) 123;
%[arXiv:hep-ph/9606251].
%%CITATION = HEP-PH 9606251;%%
M.~Drees, S.~Pakvasa, X.~Tata and T.~ter Veldhuis,
%``A supersymmetric resolution of solar and atmospheric neutrino
%puzzles,''
{\sl Phys.~Rev.} {\bf D57} (1998) 5335;
%[arXiv:hep-ph/9712392].
%%CITATION = HEP-PH 9712392;%%
E.J.~Chun, S.K.~Kang, C.W.~Kim and U.W.~Lee,
%``Supersymmetric neutrino masses and mixing with R-parity violation,''
{\sl Nucl.~Phys.} {\bf B544} (1999) 89;
%[arXiv:hep-ph/9807327].
%%CITATION = HEP-PH 9807327;%%
V.~Bednyakov, A.~Faessler and S.~Kovalenko,
%``Super-Kamiokande constraints on R-parity violating supersymmetry,''
{\sl Phys.~Lett.} {\bf B442} (1998) 203;
%[arXiv:hep-ph/9808224].
%%CITATION = HEP-PH 9808224;%%
B.~Mukhopadhyaya, S.~Roy and F.~Vissani,
%``Correlation between neutrino oscillations and collider signals of
%supersymmetry in an R-parity violating model,''
{\sl Phys.~Lett.} {\bf B443} (1998) 191;
%[arXiv:hep-ph/9808265].
%%CITATION = HEP-PH 9808265;%%
O.C.W.~Kong,
%``Neutrino oscillations and flavor structure of supersymmetry without
%R-parity,''
{\sl Mod.~Phys.~Lett.} {\bf A14} (1999) 903;
%[arXiv:hep-ph/9808304].
%%CITATION = HEP-PH 9808304;%%
J.~Ferrandis,
%``Basis independent study of supersymmetry without R-parity and the
%tau-neutrino mass,''
{\sl Phys.~Rev.} {\bf D60} (1999) 095012;
%[arXiv:hep-ph/9810371].
%%CITATION = HEP-PH 9810371;%%
D.E.~Kaplan and A.E.~Nelson,
%``Solar and atmospheric neutrino oscillations from bilinear R-parity
%violation,''
{\sl JHEP} {\bf 0001} (2000) 033;
%[arXiv:hep-ph/9901254].
%%CITATION = HEP-PH 9901254;%%
A.S.~Joshipura and S.K.~Vempati,
%``Sneutrino vacuum expectation values and neutrino anomalies through
%trilinear R-parity violation,''
{\sl Phys.~Rev.} {\bf D60} (1999) 111303;
%[arXiv:hep-ph/9903435].
%%CITATION = HEP-PH 9903435;%%
A.~Abada and M.~Losada,
%``Constraints on a general 3-generation neutrino mass matrix from
%neutrino data: Application to the MSSM with R-parity violation,''
{\sl Nucl.~Phys.} {\bf B585} (2000) 45;
%[arXiv:hep-ph/9908352].
%%CITATION = HEP-PH 9908352;%%
M.~Hirsch, M.A.~Diaz, W.~Porod, J.C.~Romao and J.W.F.~Valle,
%``Neutrino masses and mixings from supersymmetry with bilinear
%R-parity violation: A theory for solar and atmospheric neutrino
%oscillations,''
{\sl Phys.~Rev.} {\bf D62} (2000) 113008
[Erratum-ibid.\ {\bf D65} (2002) 119901];
%[arXiv:hep-ph/0004115].
%%CITATION = HEP-PH 0004115;%%
S.~Davidson and M.~Losada,
%``Basis independent neutrino masses in the R(p) violating MSSM,''
{\sl Phys.~Rev.} {\bf D65} (2002) 075025.
%[arXiv:hep-ph/0010325].
%%CITATION = HEP-PH 0010325;%%

\bibitem{lopsided}
K.S.~Babu and S.M.~Barr,
%``Large neutrino mixing angles in unified theories,''
{\sl Phys.~Lett.} {\bf B381} (1996) 202;
%[arXiv:hep-ph/9511446].
%%CITATION = HEP-PH 9511446;%%
J.~Sato and T.~Yanagida,
%``Large lepton mixing in a coset-space family unification on
%E(7)/SU(5) x U(1)**3,''
{\sl Phys.~Lett.} {\bf B430} (1998) 127;
%[arXiv:hep-ph/9710516].
%%CITATION = HEP-PH 9710516;%%
C.H.~Albright, K.S.~Babu and S.M.~Barr,
%``A minimality condition and atmospheric neutrino oscillations,''
{\sl Phys.~Rev.~Lett.} {\bf 81} (1998) 1167;
%[arXiv:hep-ph/9802314].
%%CITATION = HEP-PH 9802314;%%
J.K.~Elwood, N.~Irges and P.~Ramond,
%``Family symmetry and neutrino mixing,''
{\sl Phys.~Rev.~Lett.} {\bf 81} (1998) 5064;
%[arXiv:hep-ph/9807228].
%%CITATION = HEP-PH 9807228;%%
Y.~Nomura and T.~Yanagida,
%``Bi-maximal neutrino mixing in SO(10)(GUT),''
{\sl Phys.~Rev.} {\bf D59} (1999) 017303;
%[arXiv:hep-ph/9807325].
%%CITATION = HEP-PH 9807325;%%
N.~Haba,
%``Composite model with neutrino large mixing,''
{\sl Phys.~Rev.} {\bf D59} (1999) 035011;
%[arXiv:hep-ph/9807552].
%%CITATION = HEP-PH 9807552;%%
G.~Altarelli and F.~Feruglio,
%``A simple grand unification view of neutrino mixing and fermion mass
%matrices,''
{\sl Phys.~Lett.} {\bf B451} (1999) 388;
%[arXiv:hep-ph/9812475].
%%CITATION = HEP-PH 9812475;%%
W.~Buchmuller and T.~Yanagida,
%``Quark lepton mass hierarchies and the baryon asymmetry,''
{\sl Phys.~Lett.} {\bf B445} (1999) 399;
%[arXiv:hep-ph/9810308].
%%CITATION = HEP-PH 9810308;%%
Z.~Berezhiani and A.~Rossi,
%``Grand unified textures for neutrino and quark mixings,''
{\sl JHEP} {\bf 9903} (1999) 002;
%[arXiv:hep-ph/9811447].
%%CITATION = HEP-PH 9811447;%%
M.~Bando and T.~Kugo,
%``Neutrino masses in E(6) unification,''
{\sl Prog.~Theor.~Phys.} {\bf 101} (1999) 1313;
%[arXiv:hep-ph/9902204].
%%CITATION = HEP-PH 9902204;%%
S.~Lola and G.~G.~Ross,
%``Neutrino masses from U(1) symmetries and the Super-Kamiokande
%data,''
{\sl Nucl.~Phys.} {\bf B553} (1999) 81;
%[arXiv:hep-ph/9902283].
%%CITATION = HEP-PH 9902283;%%
Y.~Nir and Y.~Shadmi,
%``Testing Abelian flavor symmetries with neutrino parameters,''
{\sl JHEP} {\bf 9905} (1999) 023;
%[arXiv:hep-ph/9902293].
%%CITATION = HEP-PH 9902293;%%
K.~Yoshioka,
%``On fermion mass hierarchy with extra dimensions,''
{\sl Mod.~Phys.~Lett.} {\bf A15} (2000) 29;
%[arXiv:hep-ph/9904433].
%%CITATION = HEP-PH 9904433;%%
P.H.~Frampton and A.~Rasin,
%``Nonabelian discrete symmetries, fermion mass textures and large
%neutrino mixing,''
{\sl Phys.~Lett.} {\bf B478} (2000) 424;
%[arXiv:hep-ph/9910522].
%%CITATION = HEP-PH 9910522;%%
M.~Bando, T.~Kugo and K.~Yoshioka,
%``Mass matrices in E(6) unification,''
{\sl Prog.~Theor.~Phys.} {\bf 104} (2000) 211;
%[arXiv:hep-ph/0003220].
%%CITATION = HEP-PH 0003220;%%
%M.~Bando, T.~Kugo and K.~Yoshioka,
%``Novel relations between lepton and quark mixings,''
{\sl Phys.~Lett.} {\bf B483} (2000) 163;
%[arXiv:hep-ph/0003231].
%%CITATION = HEP-PH 0003231;%%
J.M.~Mira, E.~Nardi, D.A.~Restrepo and J.W.F.~Valle,
%``Bilinear R-parity violation and small neutrino masses:  A
%self-consistent framework,''
{\sl Phys.~Lett.} {\bf B492} (2000) 81;
%[arXiv:hep-ph/0007266].
%%CITATION = HEP-PH 0007266;%%
M.~Bando {\it et al.}, %T.~Kobayashi, T.~Noguchi and K.~Yoshioka,
%``Fermion mass hierarchies and small mixing angles from extra
%dimensions,'' 
{\sl Phys.~Rev.} {\bf D63} (2001) 113017;
%[arXiv:hep-ph/0008120].
%%CITATION = HEP-PH 0008120;%%
N.~Haba and H.~Murayama,
%``Anarchy and hierarchy,''
{\sl Phys.~Rev.} {\bf D63} (2001) 053010;
%[arXiv:hep-ph/0009174].
%%CITATION = HEP-PH 0009174;%%
M.S.~Berger and K.~Siyeon,
%``Anomaly-free flavor symmetry and neutrino anarchy,''
{\sl Phys.~Rev.} {\bf D63} (2001) 057302;
%[arXiv:hep-ph/0010245].
%%CITATION = HEP-PH 0010245;%%
A.~Kageyama, M.~Tanimoto and K.~Yoshioka,
%``Low-energy constraints from unification of matter multiplets,''
{\sl Phys.~Lett.} {\bf B512} (2001) 349;
%[arXiv:hep-ph/0102006].
%%CITATION = HEP-PH 0102006;%%
K.S.~Babu and S.~M.~Barr,
%``Bimaximal neutrino mixings from lopsided mass matrices,''
{\sl Phys.~Lett.} {\bf B525} (2002) 289;
%[arXiv:hep-ph/0111215].
%%CITATION = HEP-PH 0111215;%%
X.J.~Bi and Y.B.~Dai,
%``LMA solution to the solar neutrino problem and a phenomenological
%charged lepton mass matrix,''
{\sl Eur.~Phys.~J.} {\bf C27} (2003) 43;
%[arXiv:hep-ph/0204317].
%%CITATION = HEP-PH 0204317;%%
K.S.~Babu, I.~Gogoladze and K.~Wang,
%``Natural R-parity, mu-term, and fermion mass hierarchy from discrete
%gauge symmetries,''
{\sl Nucl.~Phys.} {\bf B660} (2003) 322;
%[arXiv:hep-ph/0212245].
%%CITATION = HEP-PH 0212245;%%
R.~Kitano and T.~j.~Li,
%``Flavor hierarchy in SO(10) grand unified theories via 5-dimensional
%wave-function localization,''
{\sl Phys.~Rev.} {\bf D67} (2003) 116004;
%[arXiv:hep-ph/0302073].
%%CITATION = HEP-PH 0302073;%%
T.~Asaka,
%``Lopsided mass matrices and leptogenesis in SO(10) GUT,''
{\sl Phys.~Lett.} {\bf B562} (2003) 291;
%[arXiv:hep-ph/0304124].
%%CITATION = HEP-PH 0304124;%%
K.S.~Babu, T.~Enkhbat and I.~Gogoladze,
%``Anomalous U(1) symmetry and lepton flavor violation,''
{\sl Nucl.~Phys.} {\bf B678} (2004) 233.
%[arXiv:hep-ph/0308093].
%%CITATION = HEP-PH 0308093;%%

\bibitem{STY}
J.~Sato, K.~Tobe and T.~Yanagida,
%``A constraint on Yukawa-coupling unification from lepton-flavor
%violating processes,''
{\sl Phys.~Lett.} {\bf B498} (2001) 189;
%[arXiv:hep-ph/0010348].
%%CITATION = HEP-PH 0010348;%%
J.~Sato and K.~Tobe,
%``Neutrino masses and lepton-flavor violation in supersymmetric
%models with lopsided Froggatt-Nielsen charges,''
{\sl Phys.~Rev.} {\bf D63} (2001) 116010.
%[arXiv:hep-ph/0012333].
%%CITATION = HEP-PH 0012333;%%

\bibitem{enhance}
A.Y.~Smirnov,
%``Seesaw enhancement of lepton mixing,''
{\sl Phys.~Rev.} {\bf D48} (1993) 3264;
%[arXiv:hep-ph/9304205].
%%CITATION = HEP-PH 9304205;%%
%A.Y.~Smirnov,
%``Neutrinos and Structure of the Intermediate Mass Scale,''
{\sl Nucl.~Phys.} {\bf B466} (1996) 25;
%[arXiv:hep-ph/9511239].
%%CITATION = HEP-PH 9511239;%%
M.~Tanimoto,
%``Seesaw enhancement of neutrino mixing due to the right-handed
%phases,''
{\sl Phys.~Lett.} {\bf B345} (1995) 477;
%[arXiv:hep-ph/9503318].
%%CITATION = HEP-PH 9503318;%%
M.~Bando, T.~Kugo and K.~Yoshioka,
%``Neutrino mass texture with large mixing,''
{\sl Phys.~Rev.~Lett.} {\bf 80} (1998) 3004;
%[arXiv:hep-ph/9710417].
%%CITATION = HEP-PH 9710417;%%
M.~Jezabek and Y.~Sumino,
%``Neutrino mixing and seesaw mechanism,''
{\sl Phys.~Lett.} {\bf B440} (1998) 327;
%[arXiv:hep-ph/9807310].
%%CITATION = HEP-PH 9807310;%%
M.~Bando and K.~Yoshioka,
%``Sterile neutrinos in a grand unified model,''
{\sl Prog.~Theor.~Phys.} {\bf 100} (1998) 1239;
%[arXiv:hep-ph/9806400].
%%CITATION = HEP-PH 9806400;%%
D.~Falcone,
%``Neutrino masses and mixings in a seesaw framework,''
{\sl Phys.~Rev.} {\bf D61} (2000) 097302;
%[arXiv:hep-ph/9909207].
%%CITATION = HEP-PH 9909207;%%
E.K.~Akhmedov, G.C.~Branco and M.N.~Rebelo,
%``Seesaw mechanism and structure of neutrino mass matrix,''
{\sl Phys.~Lett.} {\bf B478} (2000) 215;
%[arXiv:hep-ph/9911364].
%%CITATION = HEP-PH 9911364;%%
T.K.~Kuo, G.H.~Wu and S.W.~Mansour,
%``Mass hierarchies and the seesaw neutrino mixing,''
{\sl Phys.~Rev.} {\bf D61} (2000) 111301;
%[arXiv:hep-ph/9912366].
%%CITATION = HEP-PH 9912366;%%
A.~Datta, F.S.~Ling and P.~Ramond,
%``Correlated hierarchy, Dirac masses and large mixing angles,''
{\sl Nucl.~Phys.} {\bf B671} (2003) 383;
%[arXiv:hep-ph/0306002].
%%CITATION = HEP-PH 0306002;%%
M.~Honda, S.~Kaneko and M.~Tanimoto,
%``Seesaw enhancement of bi-large mixing in two zero textures,''
{\sl Phys.~Lett.} {\bf B593} (2004) 165;
%[arXiv:hep-ph/0401059].
%%CITATION = HEP-PH 0401059;%%
R.~Dermisek,
%``Neutrino masses and mixing, quark lepton symmetry and strong
%right-handed neutrino hierarchy,''
{\sl Phys.~Rev.} {\bf D70} (2004) 073016;
%[arXiv:hep-ph/0406017].
%%CITATION = HEP-PH 0406017;%%
M.~Lindner, M.A.~Schmidt and A.Y.~Smirnov,
%``Screening of Dirac flavor structure in the seesaw and neutrino
%mixing,''
{\sl JHEP} {\bf 0507} (2005) 048.
%[arXiv:hep-ph/0505067].
%%CITATION = HEP-PH 0505067;%%

\bibitem{flavor}
E.~Derman and H.S.~Tsao,
%``SU(2) X U(1) X S(N) Flavor Dynamics And A Bound On The Number Of
%Flavors,'' 
{\sl Phys.~Rev.} {\bf D20} (1979) 1207;
%%CITATION = PHRVA,D20,1207;%%
Y.~Yamanaka, H.~Sugawara and S.~Pakvasa,
%``Permutation Symmetries And The Fermion Mass Matrix,''
{\sl Phys.~Rev.} {\bf D25} (1982) 1895 [Erratum-ibid. {\bf D29} (1984)
2135];
%%CITATION = PHRVA,D25,1895;%%
G.~Ecker,
%``Generalized Permutation Symmetry And The Flavor Problem In SU(2)-L
%X U(1),''
{\sl Z.~Phys.} {\bf C24} (1984) 353;
%%CITATION = ZEPYA,C24,353;%%
D.~Chang, W.~Y.~Keung and G.~Senjanovic,
%``Neutrino Transitional Magnetic Moment And Nonabelian Discrete
%Symmetry,''
{\sl Phys.~Rev.} {\bf D42} (1990) 1599;
%%CITATION = PHRVA,D42,1599;%%
D.B.~Kaplan and M.~Schmaltz,
%``Flavor unification and discrete nonAbelian symmetries,''
{\sl Phys.~Rev.} {\bf D49} (1994) 3741;
%[arXiv:hep-ph/9311281].
%%CITATION = HEP-PH 9311281;%%
P.H.~Frampton and T.W.~Kephart,
%``Simple nonAbelian finite flavor groups and fermion masses,''
{\sl Int.~J.~Mod.~Phys.} {\bf A10} (1995) 4689;
%[arXiv:hep-ph/9409330].
%%CITATION = HEP-PH 9409330;%%
M.~Schmaltz,
%``Neutrino oscillations from discrete nonAbelian family symmetries,''
{\sl Phys.~Rev.} {\bf D52} (1995) 1643;
%[arXiv:hep-ph/9411383].
%%CITATION = HEP-PH 9411383;%%
P.H.~Frampton and O.C.W.~Kong,
%``Horizontal symmetry for quark and squark masses in supersymmetric
%SU(5),''
{\sl Phys.~Rev.~Lett.} {\bf 77} (1996) 1699;
%[arXiv:hep-ph/9603372].
%%CITATION = HEP-PH 9603372;%%
A.~Aranda, C.D.~Carone and R.F.~Lebed,
%``U(2) flavor physics without U(2) symmetry,''
{\sl Phys.~Lett.} {\bf B474} (2000) 170;
%[arXiv:hep-ph/9910392].
%%CITATION = HEP-PH 9910392;%%
E.~Ma and G.~Rajasekaran,
%``Softly broken A(4) symmetry for nearly degenerate neutrino masses,''
{\sl Phys.~Rev.} {\bf D64} (2001) 113012;
%[arXiv:hep-ph/0106291].
%%CITATION = HEP-PH 0106291;%%
K.S.~Babu, E.~Ma and J.W.F.~Valle,
%``Underlying A(4) symmetry for the neutrino mass matrix and the quark
%mixing matrix,''
{\sl Phys.~Lett.} {\bf B552} (2003) 207;
%[arXiv:hep-ph/0206292].
%%CITATION = HEP-PH 0206292;%%
E.~Ma,
%``Plato's fire and the neutrino mass matrix,''
{\sl Mod.~Phys.~Lett.} {\bf A17} (2002) 2361;
%[arXiv:hep-ph/0211393].
%%CITATION = HEP-PH 0211393;%%
W.~Grimus and L.~Lavoura,
%``A discrete symmetry group for maximal atmospheric neutrino mixing,''
{\sl Phys.~Lett.} {\bf B572} (2003) 189;
%[arXiv:hep-ph/0305046].
%%CITATION = HEP-PH 0305046;%%
C.I.~Low and R.R.~Volkas,
%``Tri-bimaximal mixing, discrete family symmetries, and a conjecture
%connecting the quark and lepton mixing matrices,''
{\sl Phys.~Rev.} {\bf D68} (2003) 033007;
%[arXiv:hep-ph/0305243].
%%CITATION = HEP-PH 0305243;%%
M.~Hirsch, J.C.~Romao, S.~Skadhauge, J.W.F.~Valle and A.~Villanova del
Moral,
%``Phenomenological tests of supersymmetric A(4) family symmetry model
%of neutrino mass,''
{\sl Phys.~Rev.} {\bf D69} (2004) 093006;
%[arXiv:hep-ph/0312265].
%%CITATION = HEP-PH 0312265;%%
W.~Grimus, A.S.~Joshipura, S.~Kaneko, L.~Lavoura and M.~Tanimoto,
%``Lepton mixing angle theta(13) = 0 with a horizontal symmetry D(4),''
{\sl JHEP} {\bf 0407} (2004) 078;
%[arXiv:hep-ph/0407112].
%%CITATION = HEP-PH 0407112;%%
M.~Frigerio, S.~Kaneko, E.~Ma and M.~Tanimoto,
%``Quaternion family symmetry of quarks and leptons,''
{\sl Phys.~Rev.} {\bf D71} (2005) 011901;
%[arXiv:hep-ph/0409187].
%%CITATION = HEP-PH 0409187;%%
K.S.~Babu and J.~Kubo,
%``Dihedral families of quarks, leptons and Higgses,''
{\sl Phys.~Rev.} {\bf D71} (2005) 056006;
%[arXiv:hep-ph/0411226].
%%CITATION = HEP-PH 0411226;%%
G.~Altarelli and F.~Feruglio,
%``Tri-bimaximal neutrino mixing from discrete symmetry in extra
%dimensions,''
{\sl Nucl.~Phys.} {\bf B720} (2005) 64;
%[arXiv:hep-ph/0504165].
%%CITATION = HEP-PH 0504165;%%
A.~Zee,
%``Obtaining the neutrino mixing matrix with the tetrahedral group,''
{\sl Phys.~Lett.} {\bf B630} (2005) 58.
%[arXiv:hep-ph/0508278].
%%CITATION = HEP-PH 0508278;%%

\bibitem{S3demo}
H.~Harari, H.~Haut and J.~Weyers,
%``Quark Masses And Cabibbo Angles,''
{\sl Phys.~Lett.} {\bf B78} (1978) 459;
%%CITATION = PHLTA,B78,459;%%
Y.~Koide,
%``A New View Of Quark And Lepton Mass Hierarchy,''
{\sl Phys.~Rev.} {\bf D28} (1983) 252;
%%CITATION = PHRVA,D28,252;%%
%Y.~Koide,
%``Quark Mass Matrix With Family Independent Quark Mixing,''
{\it ibid.} {\bf D39} (1989) 1391;
%%CITATION = PHRVA,D39,1391;%%
P.~Kaus and S.~Meshkov,
%``A BCS Quark Mass Matrix,''
{\sl Mod.~Phys.~Lett.} {\bf A3} (1988) 1251
[Erratum-ibid.\ {\bf A4} (1989) 603];
%%CITATION = MPLAE,A3,1251;%%
L.~Lavoura,
%``A Relationship Between The Democratic Family Mixing And The Fritzsch 
%Schemes For The Mass Matrices,''
{\sl Phys.~Lett.} {\bf B228} (1989) 245;
%%CITATION = PHLTA,B228,245;%%
M.~Tanimoto,
%``New Quark Mass Matrix Based On The Bcs Form,''
{\sl Phys.~Rev.} {\bf D41} (1990) 1586;
%%CITATION = PHRVA,D41,1586;%%
G.C.~Branco, J.I.~Silva-Marcos and M.N.~Rebelo,
%``Universal Strength For Yukawa Couplings,''
{\sl Phys.~Lett.} {\bf B237} (1990) 446;
%%CITATION = PHLTA,B237,446;%%
H.~Fritzsch and J.~Plankl,
%``Flavor Democracy And The Lepton - Quark Hierarchy,''
{\sl Phys.~Lett.} {\bf B237} (1990) 451;
%%CITATION = PHLTA,B237,451;%%
H.~Fritzsch and Z.Z.~Xing,
%``Lepton Mass Hierarchy and Neutrino Oscillations,''
{\sl Phys.~Lett.} {\bf B372} (1996) 265;
%[arXiv:hep-ph/9509389].
%%CITATION = HEP-PH 9509389;%%
M.~Fukugita, M.~Tanimoto and T.~Yanagida,
%``Atmospheric neutrino oscillation and a phenomenological lepton mass
%matrix,''
{\sl Phys.~Rev.} {\bf D57} (1998) 4429;
%[arXiv:hep-ph/9709388].
%%CITATION = HEP-PH 9709388;%%
R.N.~Mohapatra and S.~Nussinov,
%``Gauge model for bimaximal neutrino mixing,''
{\sl Phys.~Lett.} {\bf B441} (1998) 299;
%[arXiv:hep-ph/9808301];
%%CITATION = HEP-PH 9808301;%%
P.F.~Harrison and W.G.~Scott,
%``Permutation symmetry, tri-bimaximal neutrino mixing and the S3
%group characters,''
{\sl Phys.~Lett.} {\bf B557} (2003) 76;
%[arXiv:hep-ph/0302025].
%%CITATION = HEP-PH 0302025;%%
F.~Caravaglios and S.~Morisi,
%``Neutrino masses and mixings with an S(3) family permutation
%symmetry,''
arXiv:hep-ph/0503234.
%%CITATION = HEP-PH 0503234;%%

\bibitem{S3real}
S.~Pakvasa and H.~Sugawara,
%``Discrete Symmetry And Cabibbo Angle,''
{\sl Phys.~Lett.} {\bf B73} (1978) 61;
%%CITATION = PHLTA,B73,61;%%
E.~Derman,
%``Flavor Unification, Tau Decay And B Decay Within The Six Quark Six
%Lepton Weinberg-Salam Model,''
{\sl Phys.~Rev.} {\bf D19} (1979) 317;
%%CITATION = PHRVA,D19,317;%%
L.J.~Hall and H.~Murayama,
%``A Geometry of the generations,''
{\sl Phys.~Rev.~Lett.} {\bf 75} (1995) 3985;
%[arXiv:hep-ph/9508296].
%%CITATION = HEP-PH 9508296;%%
C.D.~Carone, L.J.~Hall and H.~Murayama,
%``$(S_3)~3$ flavor symmetry and $p\to K~0 e~+$,''
{\sl Phys.~Rev.} {\bf D53} (1996) 6282;
%[arXiv:hep-ph/9512399].
%%CITATION = HEP-PH 9512399;%%
R.N.~Mohapatra, A.~Perez-Lorenzana and C.A.~de Sousa Pires,
%``Type II seesaw and a gauge model for the bimaximal mixing
%explanation of neutrino puzzles,''
{\sl Phys.~Lett.} {\bf B474} (2000) 355;
%[arXiv:hep-ph/9911395].
%%CITATION = HEP-PH 9911395;%%
J.~Kubo, A.~Mondragon, M.~Mondragon and E.~Rodriguez-Jauregui,
%``The flavor symmetry,''
{\sl Prog.~Theor.~Phys.} {\bf 109} (2003) 795;
%[arXiv:hep-ph/0302196].
%%CITATION = HEP-PH 0302196;%%
T.~Kobayashi, J.~Kubo and H.~Terao,
%``Exact S(3) symmetry solving the supersymmetric flavor problem,''
{\sl Phys.~Lett.} {\bf B568} (2003) 83;
%[arXiv:hep-ph/0303084].
%%CITATION = HEP-PH 0303084;%%
J.~Kubo, H.~Okada and F.~Sakamaki,
%``Higgs potential in minimal S(3) invariant extension of the standard
%model,''
{\sl Phys.~Rev.} {\bf D70} (2004) 036007;
%[arXiv:hep-ph/0402089].
%%CITATION = HEP-PH 0402089;%%
T.~Araki, J.~Kubo and E.A.~Paschos,
%``S(3) flavor symmetry and leptogenesis,''
{\sl Eur.~Phys.~J.} {\bf C45} (2006) 465.
%[arXiv:hep-ph/0502164].
%%CITATION = HEP-PH 0502164;%%

\bibitem{S3cpx}
D.~Wyler,
%``The Cabibbo Angle In The SU(2)-L X U(1) Gauge Theories,''
{\sl Phys.~Rev.} {\bf D19} (1979) 330;
%%CITATION = PHRVA,D19,330;%%
E.~Ma,
%``Two Derivable Relationships Among Quark Masses And Mixing Angles,''
{\sl Phys.~Rev.} {\bf D43} (1991) 2761;
%%CITATION = PHRVA,D43,2761;%%
N.G.~Deshpande, M.~Gupta and P.B.~Pal,
%``Flavor changing processes and CP violation in S(3) x Z(3) model,''
{\sl Phys.~Rev.} {\bf D45} (1992) 953;
%%CITATION = PHRVA,D45,953;%%
R.~Dermisek and S.~Raby,
%``Fermion masses and neutrino oscillations in SO(10) SUSY GUT with 
%D(3) x U(1) family symmetry,''
{\sl Phys.~Rev.} {\bf D62} (2000) 015007;
%[arXiv:hep-ph/9911275].
%%CITATION = HEP-PH 9911275;%%
S.L.~Chen, M.~Frigerio and E.~Ma,
%``Large neutrino mixing and normal mass hierarchy: A discrete
%understanding,'' 
{\sl Phys.~Rev.} {\bf D70} (2004) 073008
[Erratum-ibid.\ {\bf D70} (2004) 079905];
%[arXiv:hep-ph/0404084];
%%CITATION = HEP-PH 0404084;%%
W.~Grimus and L.~Lavoura,
%``S(3) x Z(2) model for neutrino mass matrices,''
JHEP {\bf 0508} (2005) 013.
%[arXiv:hep-ph/0504153].
%%CITATION = HEP-PH 0504153;%%

\bibitem{FCNC}
S.~Dimopoulos and H.~Georgi,
%``Softly Broken Supersymmetry And SU(5),''
{\sl Nucl.~Phys.} {\bf B193} (1981) 150;
%%CITATION = NUPHA,B193,150;%%
J.R.~Ellis and D.V.~Nanopoulos,
%``Flavor Changing Neutral Interactions In Broken Supersymmetric
%Theories,''
{\sl Phys.~Lett.} {\bf B110} (1982) 44;
%%CITATION = PHLTA,B110,44;%%
J.F.~Donoghue, H.P.~Nilles and D.~Wyler,
%``Flavor Changes In Locally Supersymmetric Theories,''
{\sl Phys.~Lett.} {\bf B128} (1983) 55;
%%CITATION = PHLTA,B128,55;%%
L.J.~Hall, V.A.~Kostelecky and S.~Raby,
%``New Flavor Violations In Supergravity Models,''
{\sl Nucl.~Phys.} {\bf B267} (1986) 415;
%%CITATION = NUPHA,B267,415;%%
F.~Gabbiani and A.~Masiero,
%``Fcnc In Generalized Supersymmetric Theories,''
{\sl Nucl.~Phys.} {\bf B322} (1989) 235;
%%CITATION = NUPHA,B322,235;%%
J.S.~Hagelin, S.~Kelley and T.~Tanaka,
%``Supersymmetric flavor changing neutral currents: Exact amplitudes
%and phenomenological analysis,''
{\sl Nucl.~Phys.} {\bf B415} (1994) 293;
%%CITATION = NUPHA,B415,293;%%
F.~Gabbiani, E.~Gabrielli, A.~Masiero and L.~Silvestrini,
%``A complete analysis of FCNC and CP constraints in general SUSY
%extensions of the standard model,''
{\sl Nucl.~Phys.} {\bf B477} (1996) 321.
%[arXiv:hep-ph/9604387].
%%CITATION = HEP-PH 9604387;%%

\bibitem{gravity}
A.H.~Chamseddine, R.~Arnowitt and P.~Nath,
%``Locally Supersymmetric Grand Unification,''
{\sl Phys.~Rev.~Lett.} {\bf 49} (1982) 970;
%%CITATION = PRLTA,49,970;%%
R.~Barbieri, S.~Ferrara and C.A.~Savoy,
%``Gauge Models With Spontaneously Broken Local Supersymmetry,''
{\sl Phys.~Lett.} {\bf B119} (1982) 343;
%%CITATION = PHLTA,B119,343;%%
L.J.~Hall, J.~Lykken and S.~Weinberg,
%``Supergravity As The Messenger Of Supersymmetry Breaking,''
{\sl Phys.~Rev.} {\bf D27} (1983) 2359;
%%CITATION = PHRVA,D27,2359;%%
N.~Ohta,
%``Grand Unified Theories Based On Local Supersymmetry,''
{\sl Prog.~Theor.~Phys.} {\bf 70} (1983) 542;
%%CITATION = PTPKA,70,542;%%
H.P.~Nilles,
%``Supersymmetry, Supergravity And Particle Physics,''
{\sl Phys.~Rept.} {\bf 110} (1984) 1.
%%CITATION = PRPLC,110,1;%%

\bibitem{FN}
C.D.~Froggatt and H.B.~Nielsen,
%``Hierarchy Of Quark Masses, Cabibbo Angles And CP Violation,''
{\sl Nucl.~Phys.} {\bf B147} (1979) 277.
%%CITATION = NUPHA,B147,277;%%

\bibitem{U1}
M.~Leurer, Y.~Nir and N.~Seiberg,
%``Mass matrix models: The Sequel,''
{\sl Nucl.~Phys.} {\bf B420} (1994) 468;
%[arXiv:hep-ph/9310320].
%%CITATION = HEP-PH 9310320;%%
L.E.~Ibanez and G.G.~Ross,
%``Fermion masses and mixing angles from gauge symmetries,''
{\sl Phys.~Lett.} {\bf B332} (1994) 100;
%[arXiv:hep-ph/9403338].
%%CITATION = HEP-PH 9403338;%%
E.~Dudas, S.~Pokorski and C.A.~Savoy,
%``Yukawa matrices from a spontaneously broken Abelian symmetry,''
{\sl Phys.~Lett.} {\bf B356} (1995) 45;
%[arXiv:hep-ph/9504292].
%%CITATION = HEP-PH 9504292;%%
P.~Binetruy, S.~Lavignac and P.~Ramond,
%``Yukawa textures with an anomalous horizontal abelian symmetry,''
{\sl Nucl.~Phys.} {\bf B477} (1996) 353;
%[arXiv:hep-ph/9601243].
%%CITATION = HEP-PH 9601243;%%
R.N.~Mohapatra and A.~Riotto,
%``Supersymmetric models with anomalous U(1) mediated supersymmetry
%breaking,''
{\sl Phys.~Rev.} {\bf D55} (1997) 4262;
%[arXiv:hep-ph/9611273].
%%CITATION = HEP-PH 9611273;%%
K.~Choi, E.J.~Chun and H.D.~Kim,
%``Supersymmetry hierarchy problems and anomalous horizontal U(1)
%symmetry,''
{\sl Phys.~Lett.} {\bf B394} (1997) 89;
%[arXiv:hep-ph/9611293].
%%CITATION = HEP-PH 9611293;%%
A.E.~Nelson and D.~Wright,
%``Horizontal, anomalous U(1) symmetry for the more minimal
%supersymmetric standard model,''
{\sl Phys.~Rev.} {\bf D56} (1997) 1598;
%[arXiv:hep-ph/9702359].
%%CITATION = HEP-PH 9702359;%%
Y.~Grossman, Y.~Nir and Y.~Shadmi,
%``Large mixing and large hierarchy between neutrinos with Abelian
%flavor symmetries,''
{\sl JHEP} {\bf 9810} (1998) 007;
%[arXiv:hep-ph/9808355].
%%CITATION = HEP-PH 9808355;%%
S.F.~King,
%``Large mixing angle MSW and atmospheric neutrinos from single
%right-handed neutrino dominance and U(1) family symmetry,''
{\sl Nucl.~Phys.} {\bf B576} (2000) 85;
%[arXiv:hep-ph/9912492].
%%CITATION = HEP-PH 9912492;%%
Q.~Shafi and Z.~Tavartkiladze,
%``Anomalous flavor U(1): Predictive texture for bi-maximal neutrino
%mixing,''
{\sl Phys.~Lett.} {\bf B482} (2000) 145;
%[arXiv:hep-ph/0002150].
%%CITATION = HEP-PH 0002150;%%
N.~Maekawa,
%``Neutrino masses, anomalous U(1) gauge symmetry and doublet-triplet
%splitting,''
{\sl Prog.~Theor.~Phys.} {\bf 106} (2001) 401;
%[arXiv:hep-ph/0104200].
%%CITATION = HEP-PH 0104200;%%
N.~Maekawa and T.~Yamashita,
%``E(6) unification, doublet-triplet splitting and anomalous U(1)A
%symmetry,''
{\sl Prog.~Theor.~Phys.} {\bf 107} (2002) 1201;
%[arXiv:hep-ph/0202050].
%%CITATION = HEP-PH 0202050;%%
M.~Kakizaki and M.~Yamaguchi,
%``U(1) flavor symmetry and proton decay in supersymmetric standard
%model,''
{\sl JHEP} {\bf 0206} (2002) 032;
%[arXiv:hep-ph/0203192].
%%CITATION = HEP-PH 0203192;%%
H.K.~Dreiner, H.~Murayama and M.~Thormeier,
%``Anomalous flavor U(1)X for everything,''
{\sl Nucl.~Phys.} {\bf B729} (2005) 278;
%[arXiv:hep-ph/0312012].
%%CITATION = HEP-PH 0312012;%%
P.H.~Chankowski, K.~Kowalska, S.~Lavignac and S.~Pokorski,
%``Update on fermion mass models with an anomalous horizontal U(1)
%symmetry,''
{\sl Phys.~Rev.} {\bf D71} (2005) 055004.
%[arXiv:hep-ph/0501071].
%%CITATION = HEP-PH 0501071;%%

\bibitem{nonzero22}
D.s.~Du and Z.z.~Xing,
%``A Modified Fritzsch ansatz with additional first order
%perturbation,''
{\sl Phys.~Rev.} {\bf D48} (1993) 2349;
%%CITATION = PHRVA,D48,2349;%%
H.~Fritzsch and Z.z.~Xing,
%``A Symmetry pattern of maximal CP violation and a determination of
%the unitarity triangle,''
{\sl Phys.~Lett.} {\bf B353} (1995) 114;
%[arXiv:hep-ph/9502297].
%%CITATION = HEP-PH 9502297;%%
P.S.~Gill and M.~Gupta,
%``Fritzsch-Xing mass matrices, V(td) and CP violating phase delta,''
{\sl Phys.~Rev.} {\bf D56} (1997) 3143;
%[arXiv:hep-ph/9707445].
%%CITATION = HEP-PH 9707445;%%
J.L.~Chkareuli and C.D.~Froggatt,
%``Where does flavour mixing come from?,''
{\sl Phys.~Lett.} {\bf B450} (1999) 158;
%[arXiv:hep-ph/9812499].
%%CITATION = HEP-PH 9812499;%%
H.~Nishiura, K.~Matsuda and T.~Fukuyama,
%``Lepton and quark mass matrices,''
{\sl Phys.~Rev.} {\bf D60} (1999) 013006;
%[arXiv:hep-ph/9902385].
%%CITATION = HEP-PH 9902385;%%
D.~Falcone and F.~Tramontano,
%``Leptogenesis and neutrino parameters,''
{\sl Phys.~Rev.} {\bf D63} (2001) 073007;
%[arXiv:hep-ph/0011053].
%%CITATION = HEP-PH 0011053;%%
R.~Rosenfeld and J.L.~Rosner,
%``Hierarchy and anarchy in quark mass matrices, or can hierarchy
%tolerate anarchy?,''
{\sl Phys.~Lett.} {\bf B516} (2001) 408;
%[arXiv:hep-ph/0106335].
%%CITATION = HEP-PH 0106335;%%
W.~Buchmuller and D.~Wyler,
%``CP violation, neutrino mixing and the baryon asymmetry,''
{\sl Phys.~Lett.} {\bf B521} (2001) 291;
%[arXiv:hep-ph/0108216].
%%CITATION = HEP-PH 0108216;%%
J.L.~Chkareuli, C.D.~Froggatt and H.B.~Nielsen,
%``Minimal mixing of quarks and leptons in the SU(3) theory of flavour,''
{\sl Nucl.~Phys.} {\bf B626} (2002) 307;
%[arXiv:hep-ph/0109156].
%%CITATION = HEP-PH 0109156;%%
M.~Bando and M.~Obara,
%``Neutrino mass matrix predicted from symmetric texture,''
{\sl Prog.~Theor.~Phys.} {\bf 109} (2003) 995.
%[arXiv:hep-ph/0302034].
%%CITATION = HEP-PH 0302034;%%

\bibitem{non}
G.C.~Branco and J.I.~Silva-Marcos,
%``NonHermitian Yukawa couplings?,''
{\sl Phys.~Lett.} {\bf B331} (1994) 390;
%%CITATION = PHLTA,B331,390;%%
T.K.~Kuo, S.W.~Mansour and G.H.~Wu,
%``Triangular textures for quark mass matrices,''
{\sl Phys.~Rev.} {\bf D60} (1999) 093004;
%[arXiv:hep-ph/9907314].
%%CITATION = HEP-PH 9907314;%%
G.C.~Branco, D.~Emmanuel-Costa and R.~Gonzalez Felipe,
%``Texture zeros and weak basis transformations,''
{\sl Phys.~Lett.} {\bf B477} (2000) 147.
%[arXiv:hep-ph/9911418].
%%CITATION = HEP-PH 9911418;%%

\bibitem{UWY}
N.~Uekusa, A.~Watanabe and K.~Yoshioka,
%``Asymmetry and minimality of quark mass matrices,''
{\sl Phys.~Rev.} {\bf D71} (2005) 094024.
%[arXiv:hep-ph/0501211].
%%CITATION = HEP-PH 0501211;%%

\bibitem{AFM}
G.~Altarelli, F.~Feruglio and I.~Masina,
%``Large neutrino mixing from small quark and lepton mixings,''
{\sl Phys.~Lett.} {\bf B472} (2000) 382.
%[arXiv:hep-ph/9907532].
%%CITATION = HEP-PH 9907532;%%

\bibitem{2nuR}
R.~Kuchimanchi and R.N.~Mohapatra,
%``Bimaximal neutrino mixing from a local SU(2) horizontal symmetry,''
{\sl Phys.~Rev.} {\bf D66} (2002) 051301;
%[arXiv:hep-ph/0207110].
%%CITATION = HEP-PH 0207110;%%
T.~Endoh, S.~Kaneko, S.K.~Kang, T.~Morozumi and M.~Tanimoto,
%``CP violation in neutrino oscillation and leptogenesis,''
{\sl Phys.~Rev.~Lett.} {\bf 89} (2002) 231601;
%[arXiv:hep-ph/0209020].
%%CITATION = HEP-PH 0209020;%%
M.~Raidal and A.~Strumia,
%``Predictions of the most minimal see-saw model,''
{\sl Phys.~Lett.} {\bf B553} (2003) 72;
%[arXiv:hep-ph/0210021].
%%CITATION = HEP-PH 0210021;%%
S.F.~King,
%``Leptogenesis - MNS link in unified models with natural neutrino
%mass hierarchy,''
{\sl Phys.~Rev.} {\bf D67} (2003) 113010;
%[arXiv:hep-ph/0211228].
%%CITATION = HEP-PH 0211228;%%
S.~Raby,
%``A natural framework for bi-large neutrino mixing,''
{\sl Phys.~Lett.} {\bf B561} (2003) 119;
%[arXiv:hep-ph/0302027].
%%CITATION = HEP-PH 0302027;%%
B.~Dutta and R.N.~Mohapatra,
%``Lepton flavor violation and neutrino mixings in a 3 x 2 seesaw
%model,'' 
{\sl Phys.~Rev.} {\bf D68} (2003) 056006;
%[arXiv:hep-ph/0305059].
%%CITATION = HEP-PH 0305059;%%
V.~Barger, D.A.~Dicus, H.J.~He and T.j.~Li,
%``Structure of cosmological CP violation via neutrino seesaw,''
{\sl Phys.~Lett.} {\bf B583} (2004) 173;
%[arXiv:hep-ph/0310278].
%%CITATION = HEP-PH 0310278;%%
W.l.~Guo and Z.z.~Xing,
%``Calculable CP-violating phases in the minimal seesaw model of
%leptogenesis and neutrino mixing,''
{\sl Phys.~Lett.} {\bf B583} (2004) 163;
%[arXiv:hep-ph/0310326].
%%CITATION = HEP-PH 0310326;%%
W.~Rodejohann,
%``Hierarchical matrices in the see-saw mechanism, large neutrino
%mixing and leptogenesis,''
{\sl Eur.~Phys.~J.} {\bf C32} (2004) 235;
%[arXiv:hep-ph/0311142].
%%CITATION = HEP-PH 0311142;%%
A.~Ibarra and G.G.~Ross,
%``Neutrino phenomenology: The case of two right handed neutrinos,''
{\sl Phys.~Lett.} {\bf B591} (2004) 285;
%[arXiv:hep-ph/0312138].
%%CITATION = HEP-PH 0312138;%%
S.h.~Chang, S.K.~Kang and K.~Siyeon,
%``Minimal seesaw model with tri/bi-maximal mixing and leptogenesis,''
{\sl Phys.~Lett.} {\bf B597} (2004) 78.
%[arXiv:hep-ph/0404187].
%%CITATION = HEP-PH 0404187;%%

\bibitem{seesaw}
T.~Yanagida, in {\sl Proceedings of the Workshop on Unified Theories
and Baryon Number in the Universe}, eds.\ O.~Sawada and A.~Sugamoto
(KEK report 79-18, 1979); M.~Gell-Mann, P.~Ramond and R.~Slansky, in
{\sl Supergravity}, eds.\ P.~van~Nieuwenhuizen and D.Z.~Freedman
(North Holland, Amsterdam, 1979).

\bibitem{FGY}
P.H.~Frampton, S.L.~Glashow and T.~Yanagida,
%``Cosmological sign of neutrino CP violation,''
{\sl Phys.~Lett.} {\bf B548} (2002) 119.
%[arXiv:hep-ph/0208157].
%%CITATION = HEP-PH 0208157;%%

\bibitem{futurenuexp}
A.~Cervera, A.~Donini, M.B.~Gavela, J.J.~Gomez Cadenas, P.~Hernandez,
O.~Mena and S.~Rigolin,
%``Golden measurements at a neutrino factory,''
{\sl Nucl.~Phys.} {\bf B579} (2000) 17;
%[arXiv:hep-ph/0002108].
%%CITATION = HEP-PH 0002108;%%
P.~Huber, M.~Lindner and W.~Winter,
%``Superbeams versus neutrino factories,''
{\sl Nucl.~Phys.} {\bf B645} (2002) 3;
%[arXiv:hep-ph/0204352].
%%CITATION = HEP-PH 0204352;%%
P.~Huber, M.~Lindner, T.~Schwetz and W.~Winter,
%``Reactor neutrino experiments compared to superbeams,''
{\sl Nucl.~Phys.} {\bf B665} (2003) 487.
%[arXiv:hep-ph/0303232].
%%CITATION = HEP-PH 0303232;%%

\bibitem{nulessbb}
For a recent review, see, S.R.~Elliott and P.~Vogel,
%``Double beta decay,''
{\sl Ann.~Rev.~Nucl.~Part.~Sci.} {\bf 52} (2002) 115 [arXiv:
hep-ph/0202264].
%%CITATION = HEP-PH 0202264;%%
\end{thebibliography}
\end{document}